\documentclass[tighten, twocolumn]{aastex631}

\usepackage[T1]{fontenc}
\usepackage{ae,aecompl}
\usepackage{graphicx}	
\usepackage{amsmath}	
\usepackage{amssymb}	
\usepackage{textcomp}
\usepackage{verbatim}
\usepackage{longtable}
\usepackage{multirow}
\usepackage[bottom]{footmisc}

\usepackage{color}
\usepackage{soul}  
\hypersetup{linkcolor=purple,citecolor=blue,urlcolor=teal}

\shorttitle{Full Sky Angular Power Spectrum Estimation}
\shortauthors{Pal, Chanda \& Saha}

\graphicspath{{./}{figures/}}

\begin{document}

\title{Estimation of Full Sky Power Spectrum between Intermediate to Large Angular Scales from Partial Sky CMB Anisotropies using Artificial Neural Network}

\author[0000-0002-9502-8510]{Srikanta Pal}
\email{srikanta18@iiserb.ac.in, psrikanta357@gmail.com}

\author{Pallav Chanda}
\email{pallavchanda24@gmail.com}

\author[0000-0002-4444-1081]{Rajib Saha}
\affiliation{Department of Physics, Indian Institute of Science Education and Research Bhopal, \\ Bhopal - 462066, Madhya Pradesh, India}
\email{rajib@iiserb.ac.in}

\begin{abstract}
Reliable extraction of cosmological information from observed \textit{cosmic microwave background} (CMB) maps may require removal of strongly foreground contaminated regions from the analysis. In this article, we employ an \textit{artificial neural network} (ANN) to predict the full sky CMB angular power spectrum between intermediate to large angular scales from the partial sky spectrum obtained from masked CMB temperature anisotropy map. We use a simple ANN architecture with one hidden layer containing $895$ neurons. Using $1.2 \times 10^{5}$ training samples of full sky and corresponding partial sky CMB angular power spectra at Healpix pixel resolution parameter $N_{side} = 256$, we show that predicted spectrum by our ANN agrees well with the \textit{target} spectrum at each realization for the multipole range $2 \leq l \leq 512$. The predicted spectra are statistically unbiased and they preserve the cosmic variance accurately. Statistically, the differences between the mean predicted and underlying theoretical spectra  are within approximately $3\sigma$. Moreover, the probability densities obtained from predicted angular power spectra agree very well with those obtained from `actual' full sky CMB angular power spectra for each multipole. Interestingly, our work shows that the significant correlations in input cut-sky spectra, due to \textit{mode-mode coupling} introduced on the partial sky, are effectively removed since the ANN learns the hidden pattern between the partial sky and full sky spectra preserving the entire statistical properties. The excellent agreement of statistical properties between the predicted and the ground-truth demonstrates the importance of using artificial intelligence systems in cosmological analysis more widely.
\end{abstract}


\section{Introduction}
\textit{Cosmic microwave background} (CMB) radiation, first detected by~\cite{Penzias_1965}, originated from the so-called surface of last scattering with a temperature $\approx 2970 K$ and fills the space with a blackbody temperature of $2.72 K$ today~\citep{Fixsen_1996,Mather_1999}. This radiation contains tiny anisotropies and inhomogeneities which encode the fundamental information about the initial conditions on the metric fluctuations (induced by the quantum fluctuations of inflaton~\citep{Guth_1982} field) as well as the detailed history of physical interaction of the CMB photons with gravitation and other entities during the later development of the Universe. Accurate measurements and unambiguous usage of this \textit{gift of the nature} is therefore of absolute importance for understanding our universe reliably. Satellite-based experiments, like Planck~\citep{Planck_2020}, WMAP~\citep{Hinshaw_2013}, COBE~\citep{Bennett_1996}, and ground-based experiments, such as SPT~\citep{Hou_2014}, ACT~\citep{Sievers_2013} constrained cosmological parameters using CMB observations. Some future experiments of CMB, Echo (aka CMB-Bharat\footnote{\url{http://cmb-bharat.in/}}), CCAT-prime~\citep{Stacey_2018}, PICO~\citep{Hanany_2019}, LiteBird~\citep{Hazumi_2020} have also been proposed to meet the monumental promise the CMB provides to understand physics of the very early Universe.

Under the assumptions of Gaussianity~\citep{Guth_1982,Allen_1987,Falk_1992,Gangui_1994,Acquaviva_2003, Maldacena_2003} and statistical isotropy~\citep{Hajian_2004} of the CMB, all statistical information of the later is encoded in the two-point correlation functions which can be represented by the angular power spectrum in the harmonic space. Estimation of this spectrum is a challenging task from the observed CMB map since astrophysical emissions from the local Universes (e.g., several diffused emissions from Milky Way and localized emissions from extragalactic point sources) contaminate the primordial and weak CMB signal. Some regions of the sky may be so heavily contaminated by the local foregrounds that even after a foreground subtraction has been performed such regions must be removed entirely from the cosmological analysis since any residual foregrounds creeping in from these regions may potentially bias the cosmological interpretation. Consequently, the angular power spectrum estimated from the resulting incomplete sky loses its well-behaved statistical properties that are otherwise present in its full sky descriptions. Due to the loss of orthogonality conditions of the spherical harmonics on the partial sky, the CMB modes on all angular scales become correlated and biased low. Since cosmological theory predicts only the full sky CMB angular power spectrum, it is important to estimate the actual full sky spectrum even after any sky-mask has been applied.

In this article, we use an \textit{artificial neural network} (ANN) to achieve this purpose using $N_{side}=256$ CMB maps for the multipole range $2 \leq l \leq 512$. An earlier work~\citep{Chanda_2021} in this direction discussed this problem over only large  angular scales on the sky ($2 \leq l \leq 32$). An important advantage of our method is that, estimating the aleatoric uncertainty (due to the input masked sky spectra) for each of the ANN predicted full sky spectra and inducing these uncertainties in the predicted full sky spectra, we are able to predict the ground-truth full sky spectra at each multipole without any need to estimate the band average spectrum which we otherwise do to reduce the correlation due to \textit{mode-mode coupling} over the partial sky and to account for the lost information in presence of sky cuts. Since our ANN effectively learns to get back the lost information due to masking, it can reproduce the 'actual' full sky spectrum at each angular scale $l$ (within small random fluctuations) effectively reducing any correlations between the spectrum of different multipoles even at any given  sky realization. We demonstrate that our multipole-wise recovered spectra are statistically completely equivalent to the underlying full sky spectra by computing the correlations matrix and cosmic variance of the former. The realization specific accurate reconstruction of the ground-truth over a wide angular scale is encouraging and demands further utilization of the ANN in the analysis of CMB angular power spectrum.

In the current work, we evolve the procedure of~\cite{Chanda_2021} with some modifications.~\cite{Chanda_2021} used an ANN with two hidden layers each containing 1024 neurons to predict the full sky spectrum from the partial sky $N_{side}=16$ CMB maps. In their work, they used a \textit{concrete dropout} method~\citep{Gal_2017}, for reducing epistemic uncertainties of the ANN model. Our ANN consists of only one hidden layer with $895$ neurons and we utilize $N_{side} = 256$ partial sky CMB maps.
We employ \textit{model averaging ensemble} method~\citep{Lai_2021} in place of \textit{concrete dropout} for minimizing epistemic uncertainties of our ANN model. When applied on the Planck data~\citep{Planck4_2020}, our ANN gives excellent agreements between the predicted and inpainted full sky spectra for COMMANDER, NILC, SMICA, SEVEM foreground cleaned CMB maps\footnote{\url{https://pla.esac.esa.int/}}.

In the contemporary literature, machine learning (ML) found its applications in several fields of physics, e.g. \textit{high energy physics, particle physics, cosmology, observational astrophysics}~\citep{Olvera_2021}. An advantage of ML is that although the Metropolis-Hastings algorithm for Bayesian inference in cosmology may be computationally expensive for large volumes of data, ANN permits a decrease in the computational time~\citep{Graff_2012,Moss_2020,Hortua_2020,Gomez1_2021}). Various types of ANN were implemented by~\cite{Mancini_2022} for the purpose of parameter estimations using CMB data, which showed that the Bayesian process can speed using ANNs. In the literature~\citep{Escamilla_2020,Wang_2020,Dialektopoulos_2021,Gomez2_2021}, we also see the operation of ANNs for non-parametric reconstructions of cosmological functions.~\cite{Baccigalupi_2000} separated different types of foreground signal, such as thermal dust emissions, galactic synchrotron and radiation emitted by galaxy clusters from CMB maps using ANN.~\cite{Petroff_2020} implemented a neural network to classify the noises from the anisotropies of CMB temperatures.

On the more traditional front, there exists the classical approach of maximum-likelihood estimation. In the context of CMB,~\cite{Gorski_1994} first developed a maximum-likelihood method to estimate quadrupolar power spectrum and spectral index of primordial perturbations using orthonormal basis specially designed for analysis over a partial sky CMB map. The method was later applied on the COBE-DMR~\citep{Smoot_1991} partial sky maps by~\cite{Gorski2_1994} and~\cite{Gorski_1996}.~\cite{Gorski_1997} develop a cosmological model independent maximum-likelihood method to estimate full sky CMB angular spectrum using the COBE-DMR sky maps.~\cite{Bond_1998} use a direct and another faster but iterative maximum-likelihood method to estimate CMB angular power spectrum using COBE-DMR and Saskatoon data~\citep{Netterfield_1997}.~\cite{Wandelt_2003} develop a fast maximum-likelihood approach to estimate CMB spectrum when the telescope scanning strategy follows the so-called ring-torus pattern. Gibbs and Bayesian sampling methods~\citep{Eriksen_2004,Alsing_2015} can be used to estimate full sky CMB angular power spectrum from partial sky spectrum.~\cite{Hansen_2002} used Gabor transforms for the calculation of full sky CMB angular power spectrum from partial sky spectrum. Another well-known method to estimate full sky CMB angular power spectrum from partial sky  spectrum is the so-called pseudo-$C_{l}$ method~\citep{Peebles_1973,Wandelt_2001,Hivon_2002}. Various extensions of this method are specified in~\cite{Reinecke_2013,Elsner_2016}. In this pseudo-$C_l$ technique and other traditional approaches, the uncertainties of the estimated full sky CMB angular power spectrum are large due to sample variance that results from loss of modes for the incomplete sky coverage. Our ANN becomes very advantageous in these circumstances by avoiding such limitations since the ANN can be trained effectively to learn the mapping from the partial sky spectra to the full sky spectra preserving the cosmic variances corresponding to these full sky spectra without affecting by the sample variances.

Rest of our article is organized as follows. We represent the relation between the ensemble average of full sky and partial sky CMB angular power spectrum in section~\ref{part_to_full_cmb}. We give a basic description about ANN in section~\ref{ann}. The detailed procedures of the simulations of full sky and partial sky CMB angular power spectra are given in section~\ref{sim_full_cmb} and section~\ref{sim_part_cmb}. We discuss, in section~\ref{app_ann}, about the architecture and working procedure of the ANN used by us. We put the results of our work in section~\ref{results}. In section~\ref{full_sky_Dl_sample}, we discuss about the realization specific full sky spectra predicted by our ANN system. In section~\ref{full_sky_Dl}, we present the mean and standard deviation of ANN predicted spectra. Significance ratios of the predictions are shown in section~\ref{sign_Dl}. We describe the results of probability density of predictions in section~\ref{PD_Dl}. In section~\ref{corr_Dl}, we discuss about the correlations in the predictions of our ANN model comparing with the correlations of input partial sky spectra. We present the predictions for observed CMB maps in section~\ref{predict_real_data}. Finally, we discuss and conclude in section~\ref{conclusion}.

\section{Formalism}
\label{formalism}
\subsection{Partial sky to full sky CMB power spectrum}
\label{part_to_full_cmb}
Full sky CMB temperature anisotropies, in spherical harmonic space, can be expressed by
\begin{eqnarray}
\quad T(\theta, \phi)-T_{0} \ = \ \delta T(\theta, \phi) \ = \ \sum\limits_{l=0}^{\infty}\sum\limits_{m=-l}^{l}a_{lm}Y_{lm}(\theta, \phi) \nonumber \\ \label{delta_T}
\end{eqnarray}
where $T_{0}$ is the average temperature of CMB radiation, $T(\theta, \phi)$ represents the CMB temperature at a particular position $(\theta, \phi)$ of sky. In the right-hand side of Equation~\ref{delta_T}, $Y_{lm}(\theta, \phi)$ defines spherical harmonic functions and $a_{lm}$ are harmonic modes of the full sky anisotropies. Harmonic modes ($a_{lm}$) have ($2l+1$) degrees of freedom for a particular $l$ as the index $m$ has the range from $-l$ to $l$.

From Equation~\ref{delta_T}, harmonic modes ($a_{lm}$) can be written as
\begin{eqnarray}
\qquad a_{lm} &=& \int\limits_{\theta=0}^{\pi}\int\limits_{\phi=0}^{2\pi}\delta T(\theta, \phi)Y_{lm}^{*}(\theta ,\phi)d\Omega \label{a_lm}
\end{eqnarray}
where $d\Omega$ is the elementary solid angle and $Y_{lm}^{*}(\theta,\phi)$ defines the complex conjugate of $Y_{lm}(\theta,\phi)$.

Full sky CMB angular power spectrum, from Equation~\ref{a_lm}, is given by
\begin{eqnarray}
C_{l} & = & \frac{1}{2l+1}\sum\limits_{m=-l}^{l}|a_{lm}|^{2}. \label{C_l}
\end{eqnarray}
Equation~\ref{C_l} represents the full sky spectrum which follows the $\chi^{2}$ distribution with mean $C_{l}^{th}$ (theoretical CMB angular power spectrum) and variance $2(C_{l}^{th})^{2}/(2l+1)$. We can map the full sky anisotropies in multipoles space using the realizations of full sky spectrum. The \textit{ensemble average} of the realizations of full sky spectra agrees with theoretical spectrum. 

We generate partial sky anisotropy map applying mask on the full sky map. Operation of mask on the full sky map is nothing but the multiplication of a finite window function ($W(\theta, \phi)$) with the full sky anisotropies~\citep{Wandelt_2001}. So the partial sky anisotropies are given by
\begin{eqnarray}
\delta\tilde{T}(\theta, \phi) & = & W(\theta, \phi)\delta T(\theta, \phi). \label{partial_delta_T}
\end{eqnarray}
Window function, in terms of spherical harmonic functions, is written by
\begin{eqnarray}
W_{lm}^{l'm'} & = & \int\limits_{\theta=0}^{\pi}\int\limits_{\phi=0}^{2\pi}Y_{l'm'}(\theta, \phi)W(\theta, \phi)Y^{*}_{lm}(\theta, \phi)d\Omega. \nonumber \\ \label{window}
\end{eqnarray}
Using Equations~\ref{delta_T} \&~\ref{window} in the Equation~\ref{partial_delta_T}, the harmonic modes of the partial sky anisotropies are expressed by
\begin{eqnarray}
\tilde{a}_{lm} & = & \sum\limits_{l'=0}^{\infty}\sum\limits_{m'=-l'}^{l'}W_{lm}^{l'm'}a_{l'm'}. \label{a_l'm'}
\end{eqnarray}
The angular power spectrum of this partial sky is defined by
\begin{eqnarray}
\tilde{C}_{l} & = & \frac{1}{2l + 1}\sum\limits_{m=-l}^{l}|\tilde{a}_{lm}|^{2}. \label{partial_C_l}
\end{eqnarray}
We can develop the relation between the \textit{ensemble averages} of the full sky and corresponding partial sky spectra using Equation~\ref{C_l}, Equation~\ref{a_l'm'} and Equation~\ref{partial_C_l}. So the relation between these two \textit{ensemble averages} is given by
\begin{eqnarray}
\bigl<\tilde{C}_{l}\bigr> & = & \sum\limits_{l'=0}^{\infty}\sum\limits_{m=-l}^{l}\sum\limits_{m'=-l'}^{l'}W_{lm}^{l'm'}\left< C_{l'}\right>(W_{lm}^{l'm'})^{*} \nonumber \\ & = & \sum\limits_{l'=0}^{\infty}M_{ll'}\left< C_{l'}\right> \label{partial_to_full}
\end{eqnarray}
where the notation $\bigl<\bigr>$ defines the \textit{ensemble average} operator. In Equation~\ref{partial_to_full}, $(W_{lm}^{l'm'})^{*}$ is the complex conjugate of $W_{lm}^{l'm'}$ and $M_{ll'}$ is known as \textit{mode-mode coupling} matrix~\citep{Hivon_2002}.

Neglecting the instrumental noises at large scales, we can also define the full sky angular power spectrum in terms of corresponding partial sky spectrum, inverting Equation~\ref{partial_to_full}, by
\begin{eqnarray}
\left< C_{l}\right> & = & \sum\limits_{l'=0}^{\infty}M_{ll'}^{-1}\bigl<\tilde{C}_{l'}\bigr>. \label{full_to_partial}
\end{eqnarray}
Equation~\ref{full_to_partial} is the fundamental equation of our work. We configure our ANN system to learn this inverse of the \textit{mode-mode coupling} matrix for predicting the full sky angular power spectrum from the corresponding partial sky spectrum.
\subsection{Artificial neural network}
\label{ann}
ANN is a mathematical framework to learn the relation between input data and output data. Let us assume that the output ($y$), as a function of input ($x$), can be written as
\begin{eqnarray}
y & = & f(x). \label{input_output_rel}
\end{eqnarray}
\begin{figure}[h!]
\centering
\includegraphics[scale=0.3]{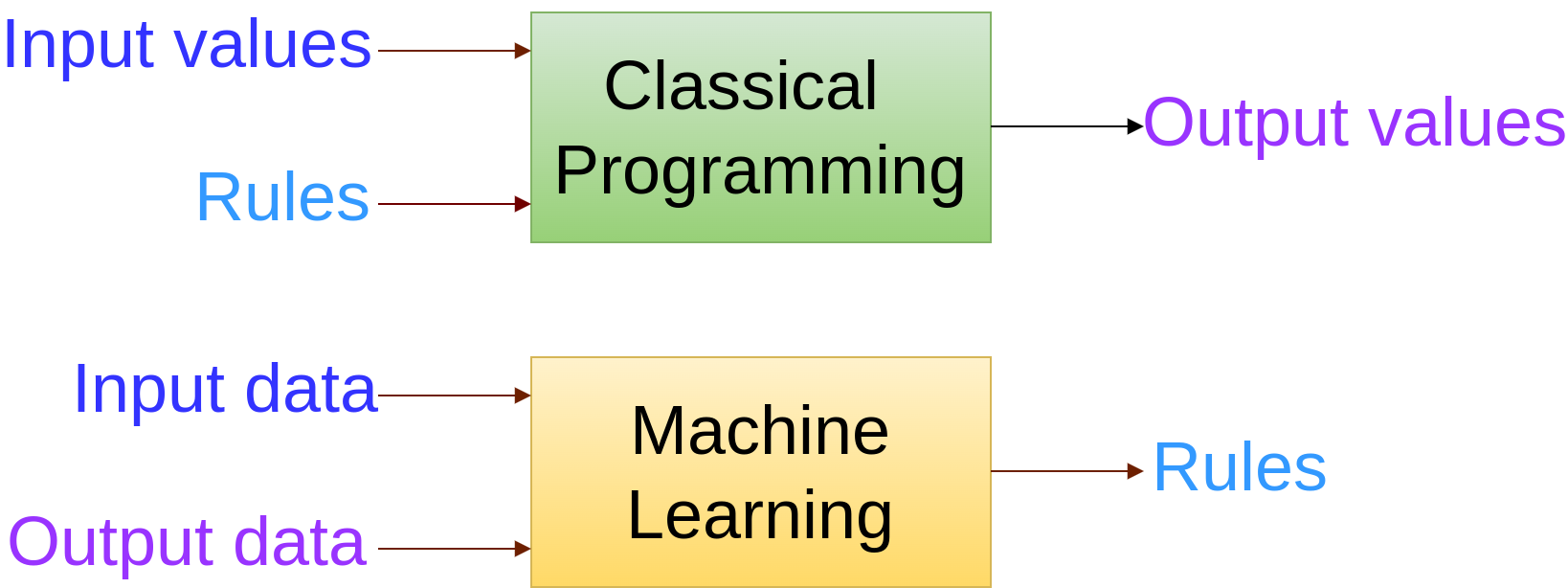}
\caption{Figure showing the block diagram of classical programming and machine learning (ML). In classical programming, output values are calculated from input values using the relation (rules) between them. ML usually learns the relation (rules) between a given set of input data and output data.}
\label{block_diagram}
\end{figure}

In classical programming, we normally calculate the value of output for a given input value using the relation (Equation~\ref{input_output_rel}) between them. In case of ML, usually, we map the relationship between a given set of input data and output data. In Figure~\ref{block_diagram}, we show the block diagram of classical programming and ML.

In our case, we employ supervised deep-learning process to predict full sky angular power spectrum using corresponding partial sky spectrum. Deep-learning, using one or more hidden layers in neural network, is a specific subfield of ML. In supervised learning, we train the ANN system using input data as well as \textit{known targets} (output data) to learn the relation between them. Common three types of ML, like \textit{binary classification, multiclass classification} and \textit{scalar} or \textit{vector regression}, are interpreted by supervised deep-learning. In present days, the widely used applications of supervised learning are such as \textit{optical character recognition, speech recognition, image classification} and \textit{language transition}. In ANN architecture, input layer contains a number of elements which are known as input features. Input layer only provides these input features to the first hidden layer. The major computations are carried by hidden layers and output layer of ANN using randomly initialized \textit{weights} and \textit{biases}.
\begin{figure}[h!]
\centering
\includegraphics[scale=0.4]{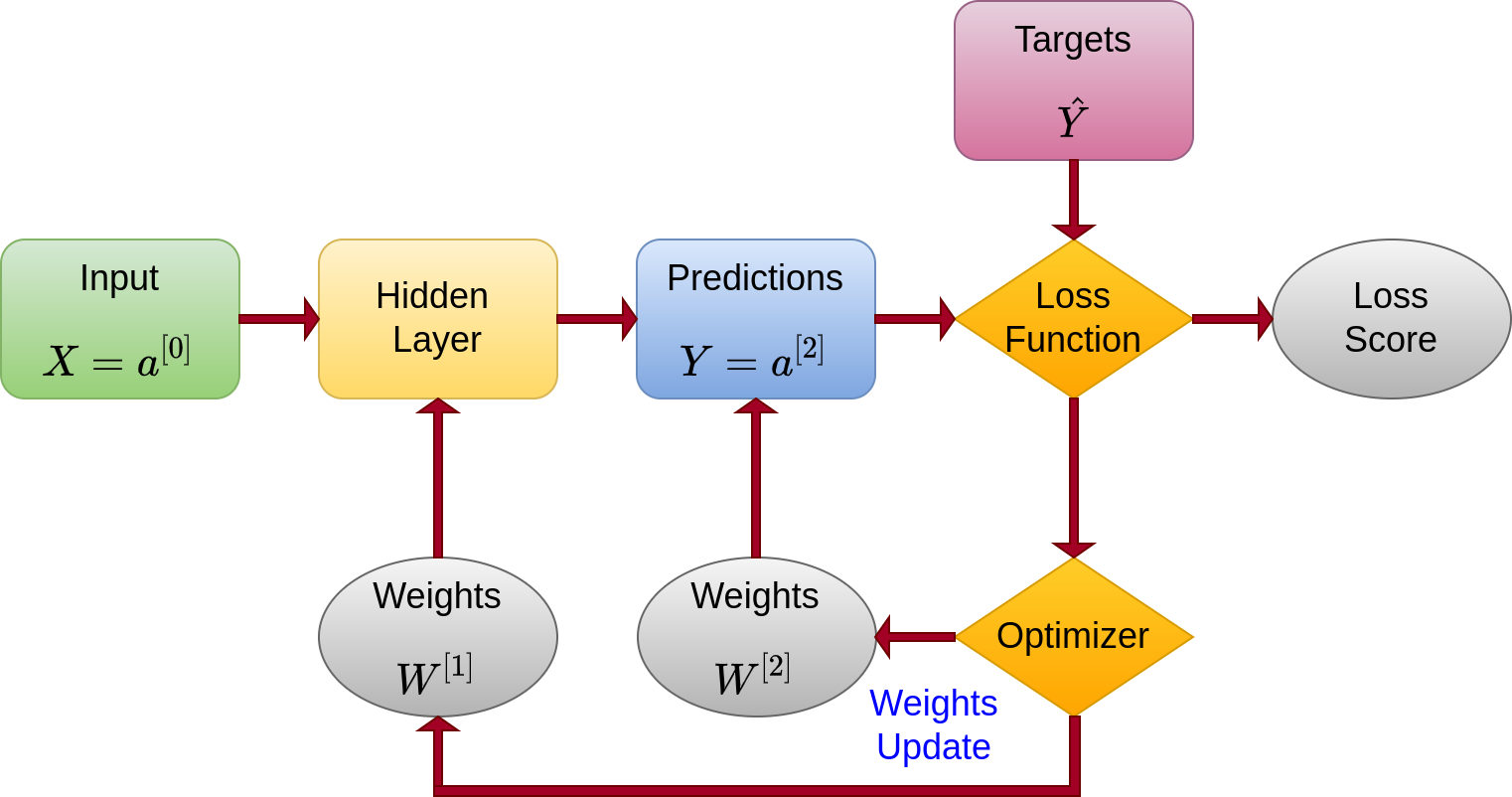}
\caption{Flowchart of \textit{forward} and \textit{backward propagation} of an \textit{artificial neural network} (ANN) with one hidden layer for \textit{vector regression}. Arrows pointing right from left follow the \textit{forward propagation} process, which takes place from input layer to \textit{loss score} estimation. \textit{Backward propagation} process is indicated by arrows pointing left from right for updating weights and \textit{optimization} processes.}
\label{flow_chart}
\end{figure}
\begin{figure}[h!]
\centering
\includegraphics[scale=0.55]{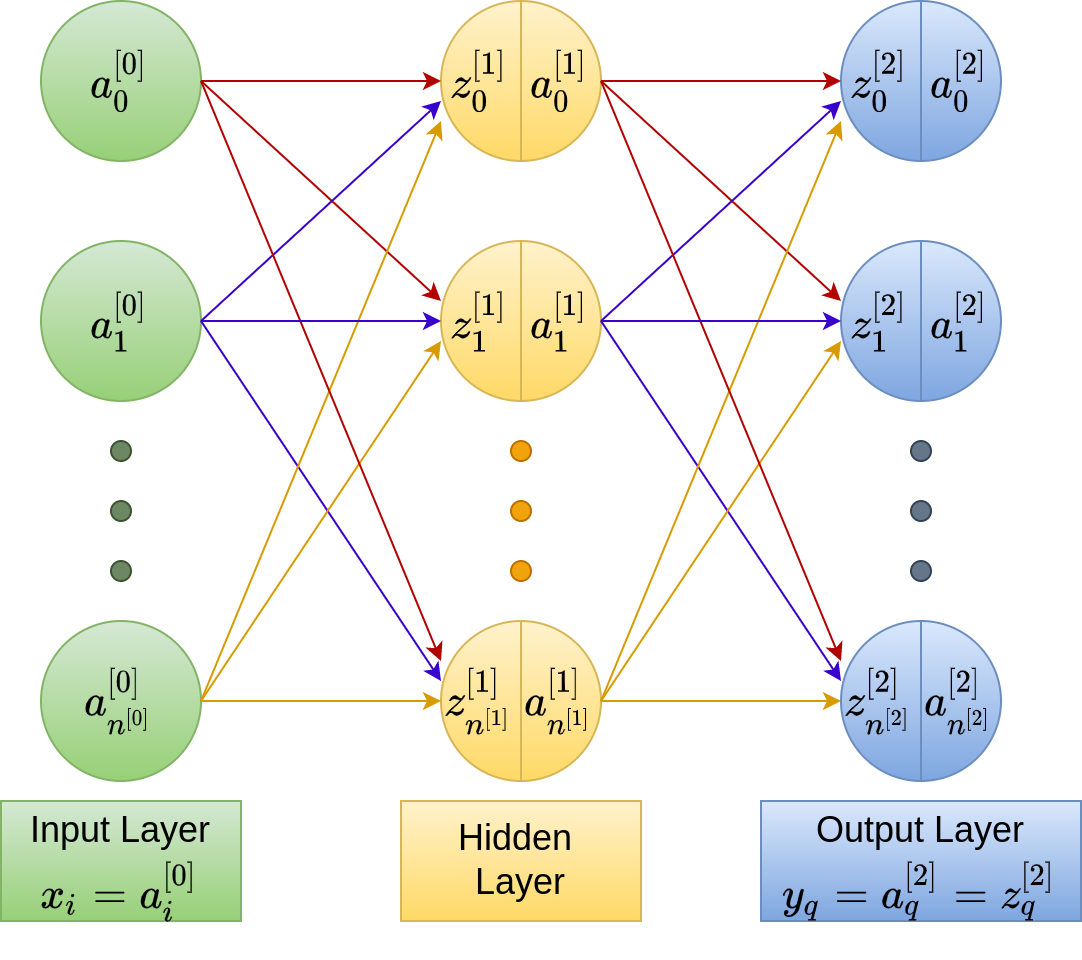}
\caption{Architecture of ANN with one hidden layer for \textit{vector regression}, where $n^{[p]}$ defines the number of neurons in $p$th layer. We show two parts in the circle (neurons of the layer) of hidden layer and output layer, where $z_{j}^{[p]}$ is linear part and $a_{j}^{[p]}$ is \textit{activation function} for $j$th neuron in $p$th layer. In input layer, $x_{i}$ are the input features given by us. In output layer, $y_{q}$ are the predictions from the ANN system.}
\label{network}
\end{figure}

ANN learns the relationship between input data and \textit{known targets} through two processes. One is \textit{forward propagation} and another is known as \textit{backward propagation}. In Figure~\ref{flow_chart}, we show a flowchart of \textit{forward} and \textit{backward propagation} processes of an ANN with one hidden layer. Let us define the input features as $\textbf{\textit{X}}$, a column matrix. Number of elements ($x_{i}$) of this column matrix is equivalent to the number of neurons in the input layer. In ANN architecture, each neuron of a particular layer connects all neurons of the previous layer by \textit{weights}. We denote the matrix representation of \textit{weights} by ${\textbf{\textit{W}}}^{[p]}$, where superscript $[p]$ defines the $p$-th layer. It is a two dimensional matrix which has the number of columns and rows same as the number of neurons in present layer and the previous layer. We define the matrix representation of \textit{biases} by ${\textbf{\textit{b}}}^{[p]}$, a column matrix, for the $p$-th layer. \textit{Bias} matrix contains the number of elements equivalent to the number of neurons of the present layer. Moreover, we need to define a \textit{activation function}, denoted by ${\textbf{\textit{a}}}^{[p]}$ for $p$-th layer, to learn the non-linearity in each layer except input layer. In input layer ($0$-th layer), ${\textbf{\textit{a}}}^{[0]}$ defines the input features ($\textbf{\textit{X}}$). \textit{Activation function} in output layer is simply a linear function for \textit{regression} problems. Now the mathematical representations of \textit{forward propagation} can be expressed by
\begin{eqnarray}
 z_{i}^{[p]} & = & \sum\limits_{j=0}^{n^{[p-1]}}w_{ij}^{[p]}a_{j}^{[p-1]} + b_{i}^{[p]} \label{z_p} \\
 a_{i}^{[p]} & = & g^{[p]}\left(z_{i}^{[p]}\right) \label{a_p}
\end{eqnarray}
where $n^{[p]}$ is the number of neurons in $p$-th layer. In Equation~\ref{z_p}, $z_{i}^{[p]}$ represents the linear part of $i$-th neuron in $p$-th layer and in Equation~\ref{a_p}, $g^{[p]}$ defines the specific \textit{activation function}. We use non-linear \textit{activation function} (e.g., \textit{sigmoid, tanh}, ReLU, LeakyReLU) in the hidden layers to learn the non-linearity between the given input and \textit{target} data set. In \textit{scalar} or \textit{vector regression} problems, we choose identity \textit{activation function} for output layer since the prediction values are real numbers. So the prediction values in the output layer can be written as
\begin{eqnarray}
y_{q} \ = \ a_{q}^{[P]} \ = \ z_{q}^{[P]} \ = \ \sum\limits_{j=0}^{n^{[P-1]}}w_{qj}^{[P]}a_{j}^{[P-1]} + b_{q}^{[P]}  \label{y_q}
\end{eqnarray}
where superscript $[P]$ specify the output layer and $y_{q}$ defines the $q$-th prediction of the output layer. We show the architecture of an ANN with one hidden layer for \textit{vector regression} problem in Figure~\ref{network}. In input layer ($p=0$), $a_{i}^{[0]}$ are the input features ($x_{i}$) of ANN, where $i=0,1,...,n^{[0]}$. In the hidden layer ($p=1$), $z_{j}^{[1]}$ are calculated by Equation~\ref{z_p} and $a_{j}^{[1]}$ are obtained from Equation~\ref{a_p}, where $j=0,1,...,n^{[1]}$. In the output layer ($P=2$), $z_{q}^{[2]}$ and $a_{q}^{[2]}$ are calculated by Equation~\ref{y_q}, where predictions $y_{q}=a_{q}^{[2]}=z_{q}^{[2]}$, for $q=0,1,...,n^{[2]}$. This is the \textit{forward propagation} part up to estimating \textit{loss score}, using \textit{loss function} (e.g., common \textit{loss function} like \textit{mean squared error} (MSE) for \textit{regression}), between predictions and \textit{known targets} in output layer. MSE \textit{loss function} can be expressed by
\begin{eqnarray}
L^{MSE} &=& \frac{1}{n^{[P]}}\sum\limits_{q=0}^{n^{[P]}}\left(y_{q}-\hat{y_{q}}\right)^{2} \label{L_mse}
\end{eqnarray}  
where $\hat{y}_{q}$ is $q$-th \textit{known target} for a particular sample and $n^{[P]}$ represents the number of neurons in the output layer. We use multiple samples of inputs ($\textbf{\textit{X}}^{k}$) and $targets$ ($\hat{\textbf{\textit{Y}}}^{k}$) to train ANN, where superscript $k$ defines $k$-th sample and $k=0,1,...,m$. The \textit{cost function} ($J^{MSE}$), for MSE \textit{loss function}, is defined by
\begin{eqnarray}
J^{MSE} &=& \frac{1}{m}\sum\limits_{k=0}^{m}L^{MSE}_{k}. \label{J_mse}
\end{eqnarray}
ANN calculates the \textit{loss score}, value of \textit{cost function}, through the \textit{forward propagation} using Equation~\ref{J_mse}. Reaching to the minimum value of \textit{loss score} signifies ANN is trained very well.

We use \textit{backward propagation} algorithm in ANN to minimize the value of \textit{cost function}~\citep{Hecht-Nielsen_1992}. In \textit{backward propagation, weights} and \textit{biases} are updated by \textit{optimization} process using a particular \textit{optimizer} (e.g., \textit{stochastic gradient descent} (SGD), \textit{momentum, adaptive moment estimation} (ADAM ;~\cite{Kingma_2014}), RMSProp~\citep{Hinton_2012}). ANN calculate the gradients of \textit{cost function} ($J$) with respect to \textit{weights} ($w_{ij}$) and \textit{biases} ($b_{i}$) through the \textit{optimization} process~\citep{Sun_2020}. These gradients can be written by
\begin{eqnarray}
\delta w_{ij} &=& \frac{\partial J}{\partial w_{ij}}. \label{delta_w} \\
\delta b_{i} &=& \frac{\partial J}{\partial b_{i}}. \label{delta_b}
\end{eqnarray}
The \textit{optimization} process utilizes the gradients of \textit{cost function}, defined in Equation~\ref{delta_w} and Equation~\ref{delta_b}, to reconstruct \textit{weights} from $w_{ij}$ to $w_{ij}-\alpha\delta w_{ij}$ and \textit{biases} from $b_{i}$ to $b_{i}-\alpha\delta b_{i}$, where $\alpha$ is \textit{learning rate} hyperparameter of the \textit{optimizer}. After updating of \textit{weights} and \textit{biases}, ANN calculates the \textit{cost function} through \textit{forward propagation} again in each iteration. These two processes help minimize the \textit{cost function} for accurate training of the ANN system. 

Estimation of uncertainties is a crucial point for data analysis in cosmology. So it will be more efficient if we are able to measure uncertainties for corresponding predictions in supervised deep-learning regression problem. We encounter two types of uncertainties in ML. One is \textit{aleatoric} uncertainty and another is \textit{epistemic} uncertainty~\citep{Kendall_2017}. \textit{Aleatoric} uncertainties appear due to the inherent noises in data. Though we can't remove these \textit{aleatoric} uncertainties from ANN, it can be measured using \textit{heteroscedastic} loss function for \textit{vector regression} problem. The \textit{heteroscedastic} (HS) loss function~\citep{Kendall_2017} can be defined by
\begin{eqnarray}
L^{HS} \ = \ \frac{1}{2n^{[p]}}\sum\limits_{q=0}^{n^{[p]}}\left[\exp\left(-s_{q}\right)\left(y_{q}-\hat{y_{q}}\right)^{2}+s_{q}\right] \label{L_hs}
\end{eqnarray}
where $s_{q}$ is the log variances $\ln\left(\sigma_{q}^{2}\right)$ and $\sigma_{q}$ is the \textit{aleatoric} uncertainties for corresponding prediction values ($y_{q}$). \textit{Epistemic} uncertainties exist in ANN due to a lack of knowledge in data as well as ignorance about model parameters. We can reduce it by taking more data samples to train ANN. We can also use one of the various techniques like \textit{model averaging ensemble}~\citep{Lai_2021}, \textit{bootstrapping, concrete dropout}~\citep{Gal_2017} to minimize \textit{epistemic} uncertainties.

\section{Methodology}
\label{method}
HEALPix\footnote{\url{https://healpix.sourceforge.io/}}~\citep{Gorski_2005} is a widely used software for simulations in cosmology. It is available in many programming languages (C, C++, IDL, Fortran, Python). We use the Python version of this software ($\texttt{healpy}$\footnote{\url{https://github.com/healpy/healpy}}) in our simulations. We take the theoretical CMB angular power spectrum ($C_{l}^{th}$), from~\cite{Planck_2020} for the purpose of simulations. We present the cosmological parameters, calculated by~\cite{Planck_2020}, in Table~\ref{cosmo_param}. We also use TensorFlow\footnote{\url{https://www.tensorflow.org/}}~\citep{Abadi_2015} ML platform to create ANN and to train that ANN for supervised deep-learning of $vector$ $regression$.
\begin{table}[h!]
\centering
\caption{Cosmological parameters obtained by~\protect\cite{Planck_2020} from standard $\Lambda$CDM model with a power law spectral index, where $\Omega_{b}$ is today's baryonic density parameter, $\Omega_{c}$ is today's density parameter of cold dark matter, $H_{0}$ is today's Hubble parameter in units of $km/s/Mpc$, $\tau$ is optical depth to decoupling surface, $n_{s}$ is scalar spectral index and $A_{s}$ is the characterize parameter for the amplitude of initial perturbations.}
\label{cosmo_param}
\begin{tabular}{lc}
\hline\hline
Parameter & Value \\ 
\hline
$\Omega_{b}h^{2}$ & $0.022 \pm 0.0001$ \\
$\Omega_{c}h^{2}$ & $0.12 \pm 0.001$ \\
$H_{0}$ & $67.37 \pm 0.54$ \\
$\Omega_{k}$ & $0.001 \pm 0.002$ \\
$\tau$ & $0.054 \pm 0.007$ \\
$n_{s}$ & $0.965 \pm 0.004$ \\
$\ln\left(10^{10}A_{s}\right)$ & $3.043 \pm 0.014$ \\ 
\hline
\end{tabular}
\end{table}
\subsection{Simulations of full sky CMB power spectra}
\label{sim_full_cmb}
\begin{figure}[h!]
\centering
\includegraphics[scale=0.34]{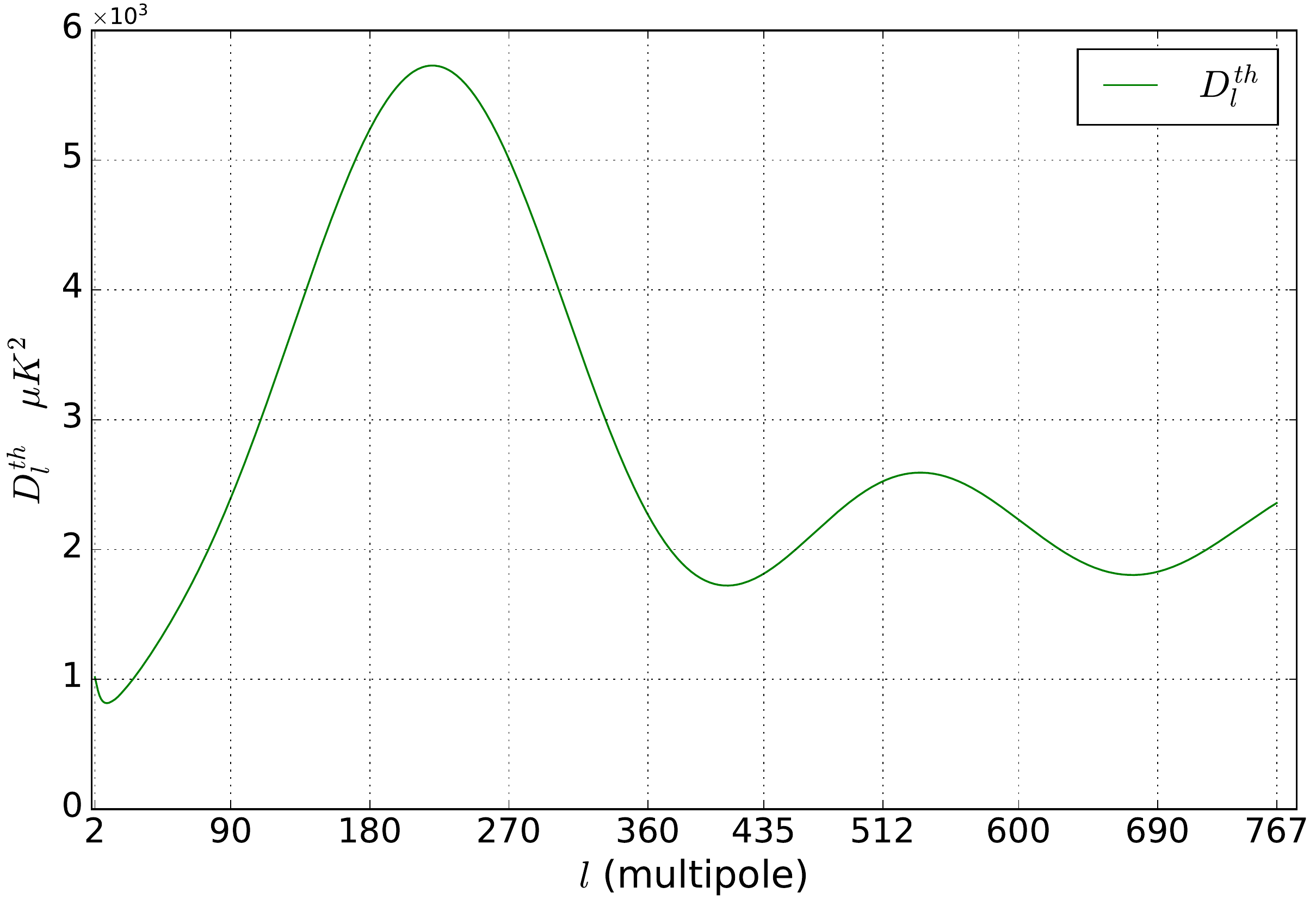}
\caption{Figure showing the theoretical angular power spectrum $D_{l}^{th}=l(l+1)C_{l}/2\pi$~\protect\citep{Planck_2020} used to simulate the random realizations of CMB maps for the multipole range $2 \leq l \leq 767$.}
\label{theory_cl}
\end{figure}
\begin{figure}[h!]
\centering
\includegraphics[scale=0.4]{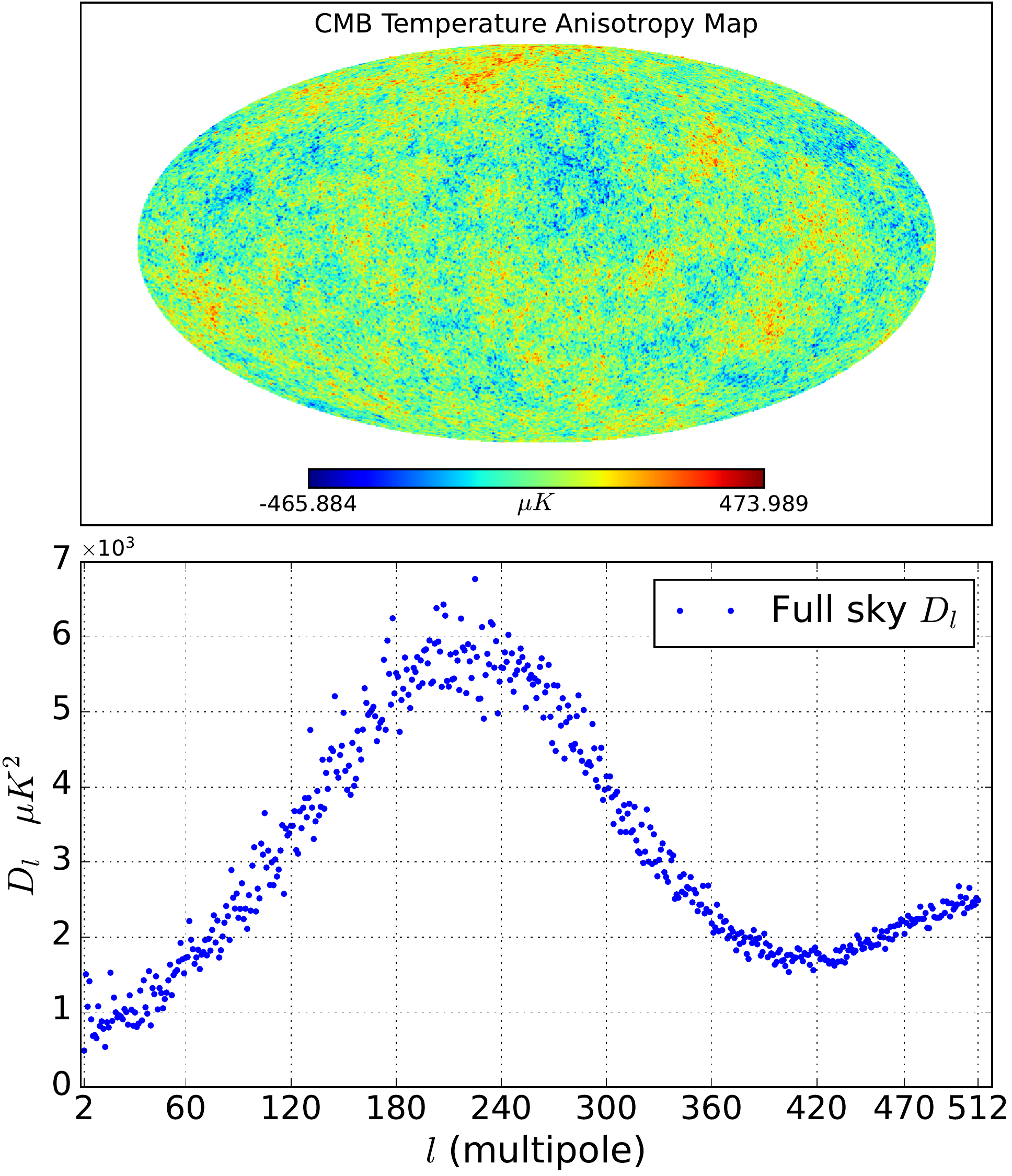}
\caption{Top panel shows the full sky CMB temperature anisotropy map, simulated from theoretical CMB angular power spectrum ($C_{l}^{th}$) with $l_{max}=767$, for a randomly selected seed value. The bottom panel represents the corresponding full sky $D_{l} = l(l+1)C_{l}/2\pi$ in $\mu K^{2}$, calculated from full sky map for $l_{max}=512$.}
\label{full_map_cl}
\end{figure}

We generate full sky CMB temperature anisotropy ($\delta T (\theta, \phi)$) map, using $\texttt{healpy.sphtfunc.synfast}$, from theoretical CMB angular power spectrum ($C_{l}^{th}$) with a maximum multipole ($l_{max}$). We work with resolution parameter $N_{side}=256$ and corresponding pixel window function ($\mathcal{P}_{l}$), provided by HEALPix software package, for the generation of full sky map. We use $l_{max} = 3N_{side}-1 = 767$ in $\texttt{healpy.sphtfunc.synfast}$ to obtain the full sky maps using randomly chosen seed values. Let us define a notation $D_{l}$ for CMB angular power spectrum multiplied by $l(l+1)/2\pi$. In Figure~\ref{theory_cl}, we show the curve of theoretical $D_{l}^{th}$ in $\mu K^{2}$ unit for the multipole range $2 \leq l \leq 767$. We ignore monopole ($l=0$) and dipole ($l=1$) power since they do not give any cosmological information. In the top panel of Figure~\ref{full_map_cl}, we show the mollweide projection of full sky map for a randomly selected seed value.

Number of pixels, for a particular $N_{side}$, can be written as
\begin{eqnarray}
N_{pix} &=& 12 \times N_{side}^{2}. \label{N_pix}
\end{eqnarray}
Angular resolution ($N_{res}$), for that specific $N_{side}$, can be expressed by
\begin{eqnarray}
N_{res} &=& \left[\sqrt{\frac{4\pi}{N_{pix}}}\frac{180}{\pi}\right]^{\circ}. \label{N_res}
\end{eqnarray}
So we can calculate, using Equation~\ref{N_pix} and Equation~\ref{N_res}, the angular resolution $13.74$ $arcmin$ for $N_{side} = 256$.

We obtain pixel-smoothed full sky spectrum ($C_{l}^{pix}$), using $\texttt{healpy.sphtfunc.anafast}$ with ring weighting and $l_{max}=2N_{side}=512$, from full sky map. For strictly band-width limited functions, $\texttt{healpy.sphtfunc.anafast}$ can easily manage the maximum multipole range $2N_{side}<l_{max}<3N_{side}-1$. We use $l_{max}=2N_{side}=512$ to generate the full sky pixel-smoothed spectrum and $l_{max}=3N_{side}-1=767$ for obtaining partial sky pixel-smoothed spectrum. These selections of maximum multipole help our ANN system to take more information, as input features, for learning the inverse of $mode$-$mode$ $coupling$ matrix beneficially. If we take $l_{max}$ as $2N_{side}$ or $3N_{side}-1$ for generating angular power spectrum for both full sky and partial sky, then ANN system will suffer from a lack of knowledge in input features to predict the full sky spectrum with same dimension as of input features.

Finally, we divide the pixel-smoothed full sky spectrum by the square of corresponding pixel window function to find the full sky spectrum ($C_{l}$). Bottom panel of Figure~\ref{full_map_cl} shows the realization of full sky $D_{l}$ in $\mu K^{2}$ unit, corresponding to the full sky map shown in the top panel of Figure~\ref{full_map_cl}, for the multipole range $2 \leq l \leq 512$. We can produce a number of realizations of full sky maps from $C_{l}^{th}$ using randomly selected seed values and generate same number of realizations of $C_{l}$ from these full sky maps. Ensemble average of these realizations of $C_{l}$ agrees with the theoretical spectrum. Using a large number of simulations helps achieve higher accuracy of the agreement by suppressing any possible residual Monte-Carlo noise in the ensemble averaged spectrum.
\subsection{Simulations of partial sky CMB power spectra}
\label{sim_part_cmb}
\begin{figure}[h!]
\centering
\includegraphics[scale=0.35]{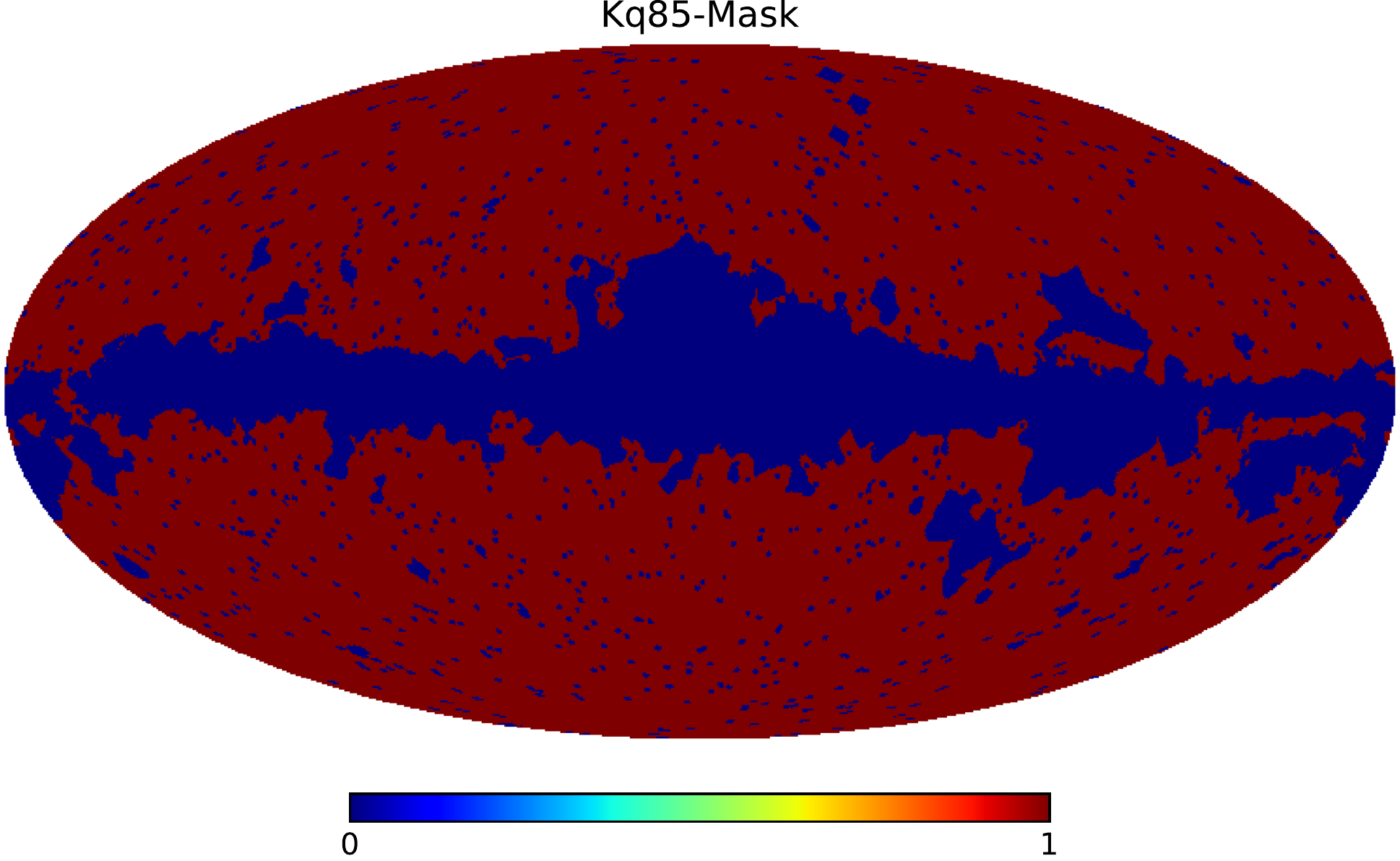}
\caption{Mollweide projection of the Kq85-mask used in this work at $N_{side} = 256$. Note that the mask removes both the galactic region and positions of the extragalactic bright point sources. For details of construction of the mask, we refer to section~\ref{sim_part_cmb}.}
\label{kp2_mask}
\end{figure}
We produce partial sky CMB temperature anisotropy map after applying WMAP Kq85-mask\footnote{\url{https://lambda.gsfc.nasa.gov/product/wmap/dr5/masks_get.html}}~\citep{Bennett_2013} on the full sky CMB map. Kq85-mask is available with pixel resolution $N_{side} = 512$. For the purpose of our analysis we downgrade it to $N_{side} = 256$.  The degradation of pixel resolution causes some pixels of the mask at lower resolution to have  fractional values between $0$ and $1$. We convert this downgraded mask to a binary mask by modifying the fractional pixel values. This is achieved by assigning all pixels of the mask originally with values larger than $0.5$ to the new value of unity. We assign the rest of the pixel values of the mask to the value $0$. We apply this binary mask on the full sky CMB maps, which excise approximately $25\%$ region from the full sky maps. Such masking is essential to remove any possible (residual) foreground contaminations arising due the galactic region and extragalactic bright point sources before cosmological analysis in case foreground cleaned CMB maps. In Figure~\ref{kp2_mask}, we show the mollweide projection of this binary Kq85-mask.

We obtain partial sky spectra ($\tilde{C}_{l}^{pix}$) using $\texttt{healpy.sphtfunc.anafast}$ on the input partial sky maps obtained above. We use ring weighting with $l_{max}=3N_{side} -1=767$ for the generation of $\tilde{C}_{l}^{pix}$ which will be served as input to the ANN. This helps to provide the ANN with as much spectral information as possible within the limits of numerical algorithms used to perform spherical harmonic transformations.  We note that these partial sky spectra contain pixel smoothing effects since the input full sky CMB maps are smoothed by the pixel window function.
\begin{figure}[h!]
\centering
\includegraphics[scale=0.4]{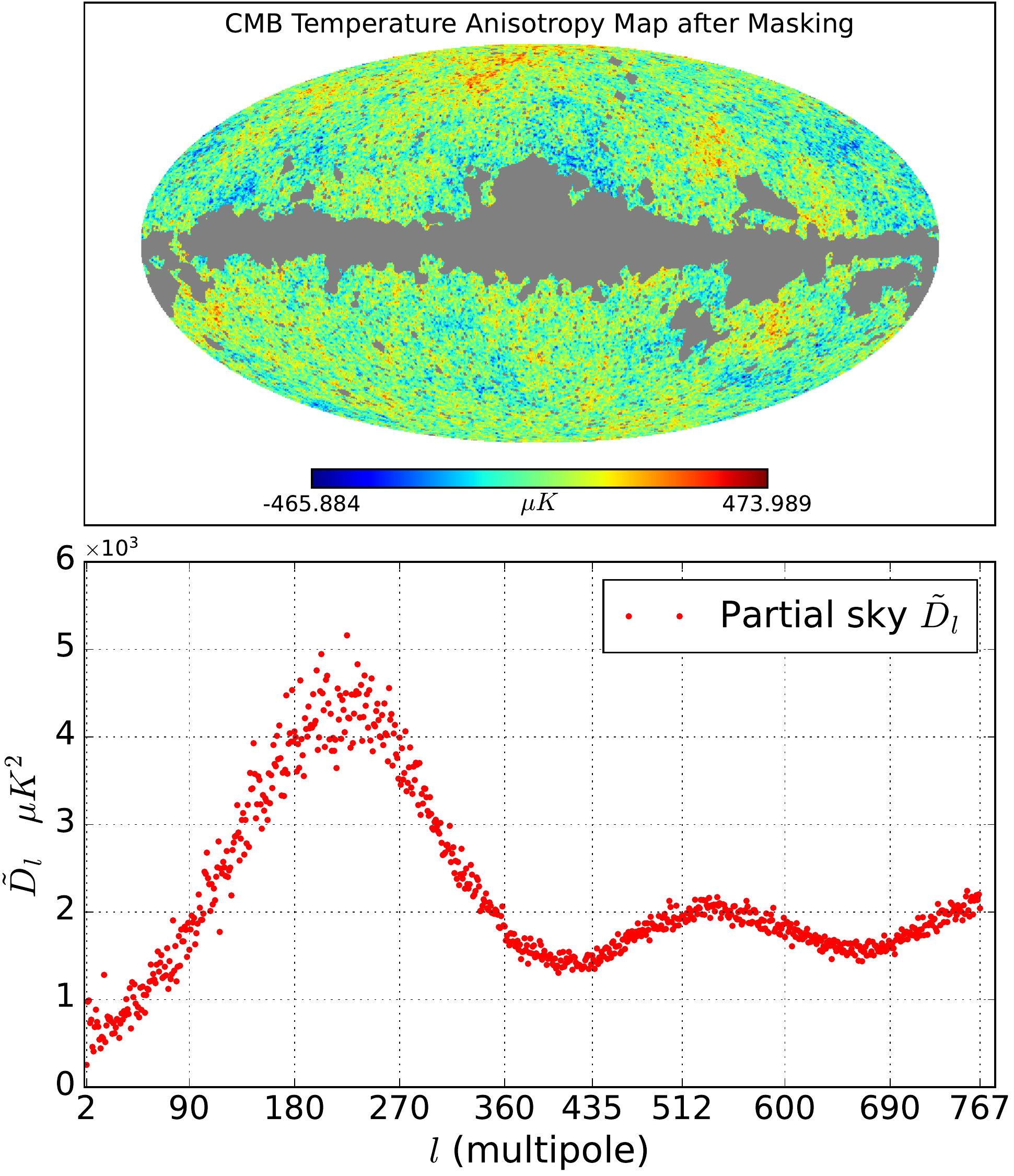}
\caption{Top panel shows the partial sky CMB temperature anisotropy map obtained by masking the full sky simulated CMB map using theoretical CMB angular power spectrum ($C_{l}^{th}$), given by~\protect\cite{Planck_2020}, with $l_{max}=767$ using a randomly chosen seed value. Bottom panel shows the corresponding partial sky $\tilde{D}_{l}=l(l+1)\tilde{C}_{l}/2\pi$ in $\mu K^{2}$ unit, calculated from partial sky map of the top panel, for the multipole range $2 \leq l \leq 767$. Reduction of power due to masking can be concluded by comparing this figure with the bottom panel of Figure~\ref{full_map_cl}.}
\label{partial_map_cl}
\end{figure}

Finally, we scale the smoothed partial sky spectrum ($\tilde{C}_{l}^{pix}$) by dividing by the square of the pixel window function to obtain a scaled spectrum ($\tilde{C}_{l}$).  We generate $1.5 \times 10^{5}$ number of realizations of partial sky $\tilde{D}_{l}=l(l+1)\tilde{C}_{l}/2\pi$. In Figure~\ref{partial_map_cl}, top panel shows the partial sky CMB temperature anisotropy map for a randomly chosen seed value. The bottom panel represents the corresponding partial sky $\tilde{D}_{l}$ in $\mu K^{2}$ unit for the multipole range $2 \leq l \leq 767$. Comparing the bottom panels of Figure~\ref{full_map_cl} and Figure~\ref{partial_map_cl}, we notice that the power of the partial sky spectrum is less compared to the full sky case. This is expected since application of mask removes some fractions of independent numbers of modes at each angular scales.
\subsection{ANN for our analysis}
\label{app_ann}
\begin{figure}[h!]
\centering
\includegraphics[scale=0.4]{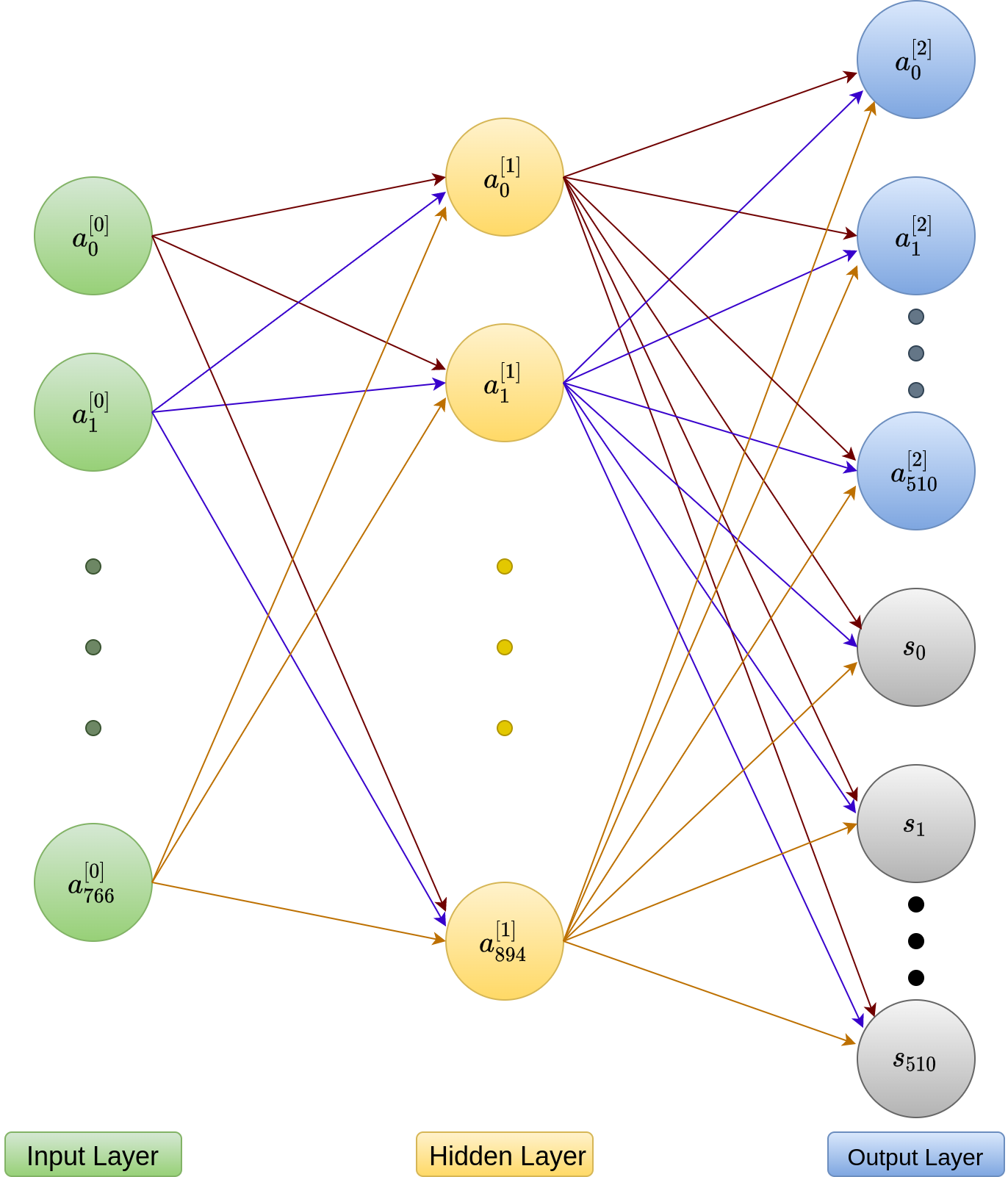}
\caption{Architecture of our ANN system with one hidden layer for \textit{heteroscedastic} loss function. In the input layer, $a_{i}^{[0]}$, where $i=0,1,...,766$, represents the input features. In the hidden layer, $a_{i}^{[1]}$, where $i=0,1,..,894$, are the values of ReLU activation function. First half of output layer represents the predictions ($y_{q}$) and second half of the output layer represents the log variances ($s_{q}$) of the predictions ($y_{q}$), where $q=0,1,...,510$.}
\label{adv_net}
\end{figure}

We use \textit{sequential} model and \textit{dense} layer (densely connected layer) from \textit{keras} library of TensorFlow ML platform in our supervised deep-learning \textit{vector regression} problem. We create an ANN with input layer, one hidden layer (contained ReLU \textit{activation} function) and output layer (carried identity \textit{activation} function). We utilize $\texttt{L2}$ \textit{kernel regularizer}, with a factor $10^{-3}$, in the hidden layer to avoid any possible overfitting in our ANN system. We use the realizations of partial sky $\tilde{D}_{l}$, for the multipole range $1 \leq l \leq 767$, as input data $\textbf{\textit{X}}^{k}$ with matrix elements $x_{i}^{k}$, where $i=0,1,...,766$ and superscript $k$ defines the $k$-th sample. We take the realizations of full sky $D_{l}$, for the multipole range $2 \leq l \leq 512$, as known targets $\hat{\textbf{\textit{Y}}}^{k}$ with matrix elements $y_{q}^{k}$, where $q=0,1,...,510$ and superscript $k$ represents the $k$-th sample. We create $1.5 \times 10^{5}$ number of samples of full sky $D_{l}$ and partial sky $\tilde{D}_{l}$. We use first $80\%$ samples of the set for training. From the rest of the samples, we use first half for validation and last half for testing the performance of our ANN system. We make use of \textit{heteroscedastic} loss function ($L^{HS}$), defined in Equation~\ref{L_hs}, in our ANN system for predicting output values as well as corresponding \textit{aleatoric} uncertainties. So the number of neurons in input layer ($n^{[0]}$) is equal to $767$ and the number of neurons in output layer ($n^{[2]}$) is equivalent to $1022$. The first half of the output layer ($0 \leq q \leq 510$) determines the predictions and last half of the output layer ($511 \leq q \leq 1021$) gives the log variances ($s_{q}$) of corresponding predictions. The number of neurons in the hidden layer ($n^{[1]}$) is calculated by a simple average of number of neurons in the input and the output layers. We take this number to be $\approx 895$ for our case. In Figure~\ref{adv_net}, we show a schematic diagram of our ANN system contained one hidden layer as well as \textit{heteroscedastic} loss function ($L^{HS}$).

Preprocessing of input data is a well-known procedure in ML. This procedure arranges the input data with a better scaling to train an ANN effectively. We perform the preprocessing using \textit{standardization} method for scaling the input data $\textbf{\textit{X}}^{k}$ with mean $0$ and standard deviation $1$. We take the mean as well as the standard deviation of the training samples. Then, each sample is subtracted by the mean and divided by the standard deviation. This \textit{preprocessing} is also applied to the validation set and test set. We standardize the validation and test sets by using the mean and standard deviations of samples obtained from the training set. This helps to effectively pass on the information learnt by the ANN during the training process for predictions in the validation and test cases. We also scale the known targets $\hat{\textbf{\textit{Y}}}^{k}$, dividing by $10^{3}$, to decrease the range of values of $\hat{\textbf{\textit{Y}}}^{k}$. This helps the ANN system to learn easily and quickly.

A common problem of an ANN system is that it can be trapped at a local minimum of the loss function (in our case it is $L^{HS}$). We employ mini-batch \textit{optimization} algorithm (e.g., \textit{mini-batch stochastic gradient descent} (MSGD), ADAM etc.) to overcome this local minima problem~\citep{Ruder_2016}. We operate our ANN system with ADAM \textit{optimizer}, where we set \textit{learning rate} hyperparameter $\alpha=10^{-4}$ for tuning our ANN system. In mini-batch \textit{optimization} algorithm, ANN system uses a subset from entire training set for each iteration. Completing all iterations, ANN is trained with the entire training set. To decide how many times the \textit{optimization} process should go on we use 200 \textit{epochs}. We also fix the mini-batch size ($m_{b}$) conveniently to a value of $2048$ which is less than the number of entire training samples ($m$). So the number of iterations, in each epoch, will be $m/m_{b}=59$.

After training our ANN system with training samples, we predict the full sky $D_{l}$, for the multipole range $2 \leq l \leq 512$, using input data ($\tilde{D}_{l}$) of $1.5 \times 10^4$ test samples. As we use \textit{heteroscedastic} loss function ($L^{HS}$), we get predictions ($y_{q}$) as well as log variances ($s_{q}$) of corresponding predictions from the same ANN system. We calculate aleatoric uncertainties ($\sigma_{q}$) taking the square root of $\exp(s_{q})$. We generate random Gaussian realizations from the Gaussian distribution with $0$ mean value and $\sigma_{q}$ as the standard deviation~\citep{Chanda_2021}. So the final results of the full sky $D_{l}$ are the sum of predicted full sky $D_{l}$, for the multipole range $2 \leq l \leq 512$, and these Gaussian realizations. ~\cite{Chanda_2021} used \textit{concrete dropout}~\citep{Gal_2017} to minimize the \textit{epistemic} uncertainties of their ANN system for $N_{side}=16$ case. We note in passing that using \textit{model averaging ensemble} method as used in this work  requires less computational resources compared with the alternative approach of computing epistemic uncertainties using the so-called \textit{concrete dropout} of the neurons. 

Here, we discuss the ability of \textit{model averaging ensemble} method~\citep{Lai_2021} to minimize the \textit{epistemic} uncertainties of our ANN system to avoid the computational cost. For this method, we concentrate on the random initialization of \textit{weights} ($\textbf{\textit{W}}^{[p]}$) and \textit{biases} ($\textbf{\textit{b}}^{[p]}$) in $dense$ layer. Output of ANN will change depending upon the random initialization of $\textbf{\textit{W}}^{[p]}$ and $\textbf{\textit{b}}^{[p]}$. We train the same ANN system, with same hyperparameter and same tuning, for a total of $50$ times for $50$ randomly chosen seed values using TensorFlow library. Then, we find the final predictions ($y_{q}^{mean}$) taking the simple mean of these $50$ output sets. For the estimation of corresponding \textit{aleatoric} uncertainties ($\sigma_{q}^{rms}$; $rms$ stands for root mean square), at first we calculate exponential of log variances ($s_{q}$) for each of $50$ output sets. Thereafter, we take the simple mean of $\exp(s_{q})$ from these $50$ output sets, and the final \textit{aleatoric} uncertainties ($\sigma_{q}^{rms}$) are obtained by taking square root of the mean values of $\exp(s_{q})$. Finally, we obtain the full sky $D_{l}$ by computing the sum of mean predictions ($y_{q}^{mean}$) and the random Gaussian realizations created from Gaussian distributions with $0$ mean and $\sigma_{q}^{rms}$ standard deviation. Thus, we get accurate results from our ANN system minimizing the \textit{epistemic} uncertainties. The entire set of $50$ iterations of training of our ANN system were performed using $\texttt{Google Colab}$\footnote{\url{https://colab.research.google.com/?utm_source=scs-index}}, a platform for an efficient online GPU service offered by Google for the purpose of ML. It takes approximately $2$ hours $15$ minutes to perform the entire set of training.

\section{Results}
\label{results}
\subsection{Realization specific predicted full sky $D_{l}$}
\label{full_sky_Dl_sample}
\begin{figure*}
\centering
\includegraphics[scale=0.45]{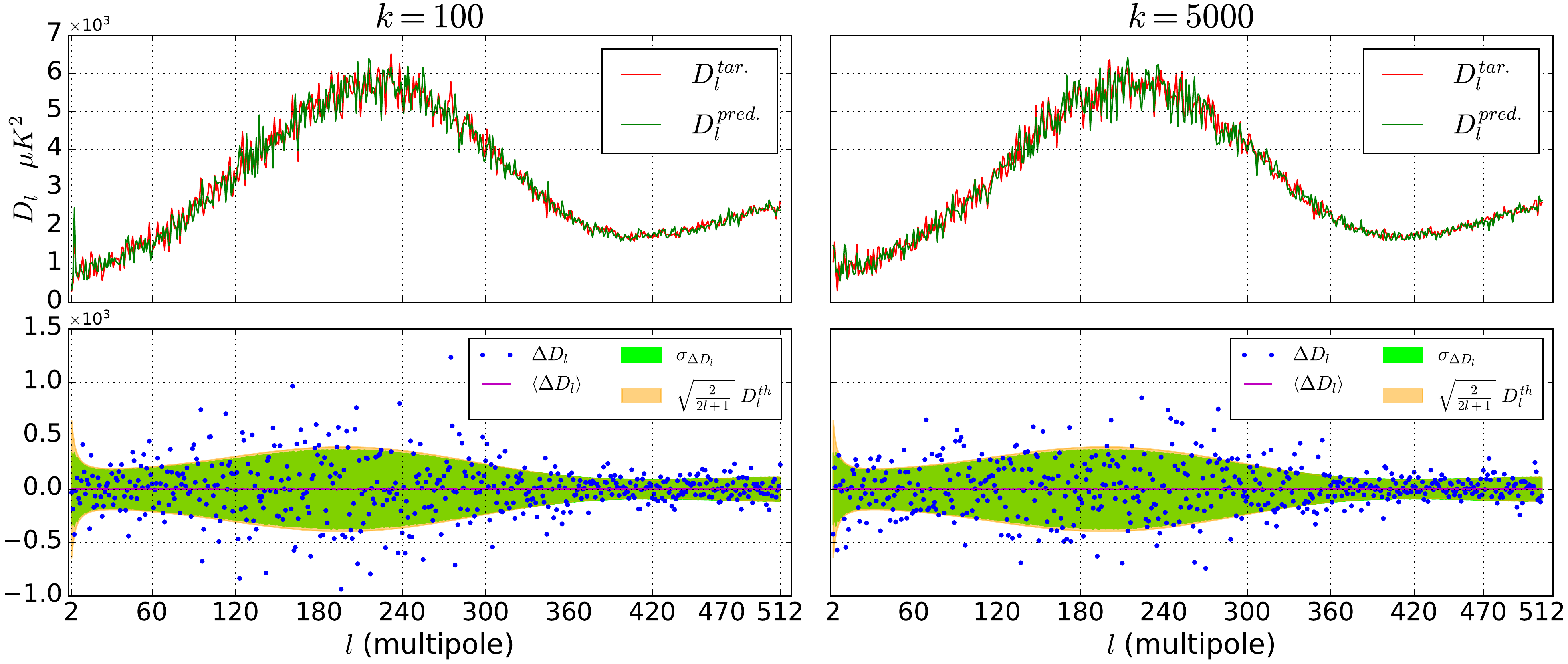}
\caption{In this figure, we present the predicted results corresponding to two randomly selected realizations ($k=100,5000$) in two columns respectively. Top panel of each column shows the agreement between the predicted (green) and target (red) full sky spectra. Bottom panel of each column shows that most of the differences ($\Delta D_l=D_l^{tar.}-D_l^{pred.}$; blue) between target and predicted spectra are within the standard deviations ($\sigma_{\Delta D_l}$; green) corresponding to these differences in case of every realization, which indicates our ANN system predicts the full sky spectrum from cut-sky map reliably. We present the standard deviation region (green) of these differences as the region in between $\left<\Delta D_l\right>-\sigma_{\Delta D_l}$ and $\left<\Delta D_l\right>+\sigma_{\Delta D_l}$, where $\left<\Delta D_l\right>$ (magenta) is the mean of the realizations of those differences. We further show the cosmic standard deviation ($\sqrt{2/(2l+1)}D_{l}^{th}$; orange) region as the region in between $-\sqrt{2/(2l+1)}D_{l}^{th}$ and $\sqrt{2/(2l+1)}D_{l}^{th}$. See section~\ref{full_sky_Dl_sample} for the detailed discussion of this figure.}
\label{test_visual}
\end{figure*}
We use a total of $1.5 \times 10^{4}$ test samples of the partial sky CMB spectra to predict the corresponding full sky spectrum for each of the test cases. We predict the realizations of the full sky $D_{l}$ for the multipole range $2 \leq l \leq 512$ using our trained ANN architecture. We show the predicted and target full sky $D_{l}$ in Figure~\ref{test_visual} for two different randomly chosen test samples ($k=100,5000$). Both the predicted and target spectra agree very well with each other within a small level of random fluctuations in each of the sub-figures. In this figure, we additionally show the differences ($\Delta D_l = D_l^{tar.}-D_l^{pred.}$) between the target and predicted spectra for these two randomly chosen realizations immediately below each power spectrum plot. In these bottom panels we also show the errorbars by computing the standard deviations ($\sigma_{\Delta D_l}$) of the differences ($\Delta D_l$) between the target and predicted spectra. These errorbars indicate the significant overlap of the data between the target and predicted spectra since the ANN makes the predictions based upon the available unmasked sky region. In these two bottom panels containing the errorbars, we also show the cosmic variance induced standard deviations (orange color). As seen from these figures, the errorbars (green) corresponding to predicted spectra are almost comparable with the cosmic variance induced errors for the input partial sky spectra produced by using Kq85-mask and most of the differences $\Delta D_l$ are within the corresponding standard deviation region (green) of these $\Delta D_l$. From these results, we conclude that the predicted spectra agree statistically with the target spectra.
\subsection{Predicted $\bigl<D_{l}\bigr>$ and $\sigma_{D_{l}}$}
\label{full_sky_Dl}
\begin{figure*}
\centering
\includegraphics[scale=0.45]{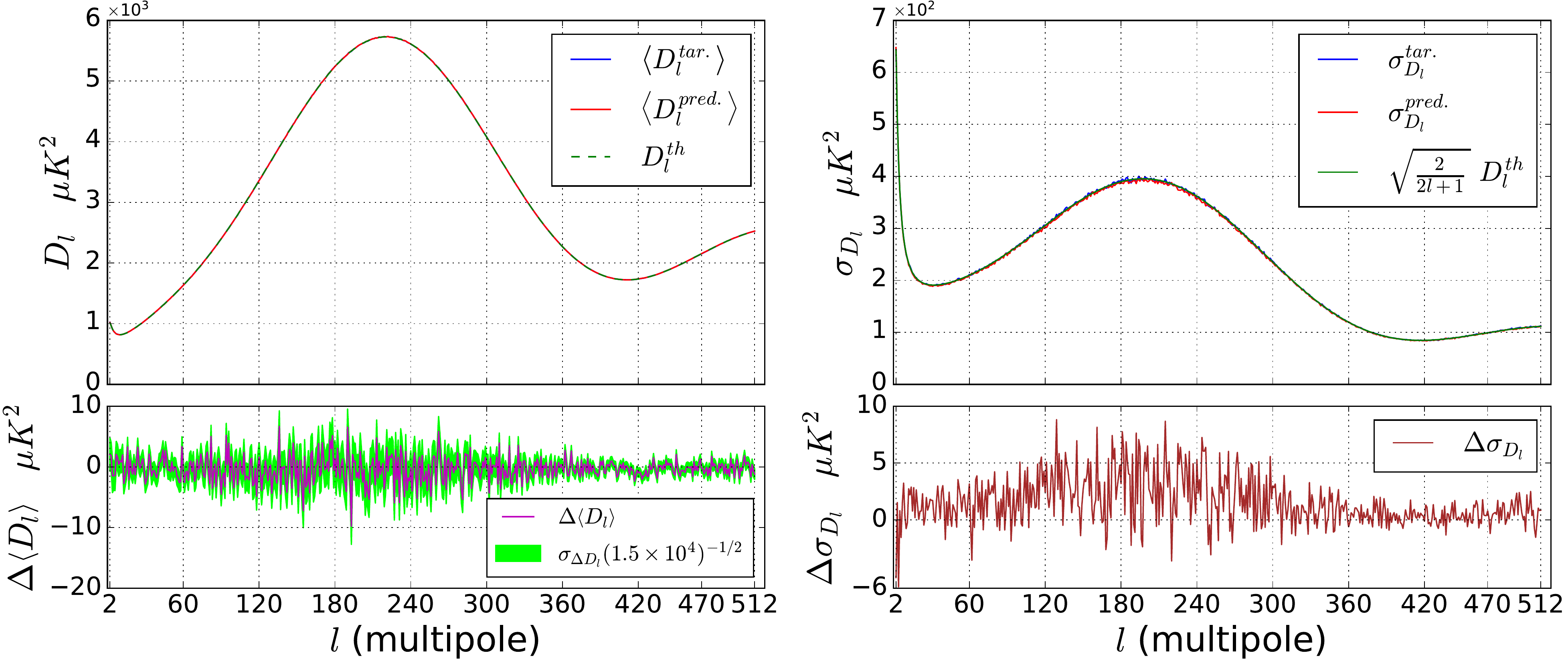}
\caption{Top panel of the left column shows the mean target full sky spectrum ($\bigl<D_{l}^{tar.}\bigr>$; blue), the mean predicted full sky spectrum($\bigl<D_{l}^{pred.}\bigr>$; red) and the theoretical spectrum ($\bigl<D_{l}^{th}\bigr>$; green dashed line) for the multipole range $2 \leq l \leq 512$. We use the notation $D_{l}=l(l+1)C_{l}/2\pi$. The difference ($\Delta \bigl<D_{l}\bigr>=\bigl<D_{l}^{tar.}\bigr>-\bigl<D_{l}^{pred.}\bigr>$) between the mean target and predicted spectra with corresponding errorbar (lime) is shown at each multipole in the bottom panel of the left column, which indicates the better agreement between the mean target and predicted spectra. In the top panel of right column, we show the standard deviations corresponding to the target spectra ($\sigma_{D_{l}}^{tar.}$; blue) and predicted spectra ($\sigma_{D_{l}}^{pred.}$; red) for the same multipole range. In this panel, we also show the cosmic standard deviation (green) fo the theoretical spectrum. In the bottom panel of right column, we show the difference ($\Delta \sigma_{D_{l}}=\sigma_{D_{l}}^{tar.}-\sigma_{D_{l}}^{pred.}$) between target and predicted standard deviation at each multipole. See the text of section~\ref{full_sky_Dl} for the detailed discussion about all panels.}
\label{mean_std_cl}
\end{figure*}
We assess any possible bias in the samples of predicted full sky spectrum by computing of the mean spectrum for the predicted and target sets of $D_{l}$ and the standard deviation corresponding to the mean spectrum. In Figure~\ref{mean_std_cl}, we show the results obtained from our ANN system.

The top panel of the left column of this figure shows the excellent agreement between the mean target ($\bigl<D_{l}^{tar.}\bigr>$; blue) and predicted ($\bigl<D_{l}^{pred.}\bigr>$; red) spectra for the multipole range $2 \leq l \leq 512$ for our ANN method. In this left-top panel, we also present the theoretical spectrum ($\bigl<D_{l}^{th}\bigr>$; green dashed line) for the multipole range $2 \leq l \leq 512$. In the bottom panel of the left column of this figure, we represent the difference ($\Delta \bigl<D_{l}\bigr>=\bigl<D_{l}^{tar.}\bigr>-\bigl<D_{l}^{pred.}\bigr>$; magenta) between mean target and predicted spectra with corresponding standard error of mean (SEM; lime) at each multipole for our ANN analysis. These differences and the corresponding errorbars are significantly low.

In Figure~\ref{mean_std_cl}, the top panel of the right column shows that the standard deviation ($\sigma_{D_l}^{pred.}$; red) corresponding to predicted spectra agrees excellently with the standard deviation ($\sigma_{D_l}^{tar.}$; blue) of target spectra as well as with the cosmic standard deviation ($\sqrt{2/(2l+1)}D_{l}^{th}$; green) in our ANN analysis. This agreement between target and predicted standard deviations is the main achievment of our ANN system. In the bottom panel of the right column of this figure, we show the difference between the standard deviations corresponding to the target and predicted spectra at each multipole, which are also significantly low. However, we note that most of the differences between the standard deviations are positive, since the standard deviations of the predicted spectra are little lower than the same of the target spectra at most of the multipoles, even after adding the aleatoric uncertainty distribution to the ANN predicted spectra. These positive differences arise due to the presence of approximately $3\%$ maximum epistemic uncertainty in the predictions of our ANN system even after applying the \textit{model averaging ensemble} method in our analysis to reduce the epistemic uncertainty.

From the results of Figure~\ref{mean_std_cl}, we conclude that the mean predicted and mean target spectra agree very well with each other for our ANN system. Also, our method recovers the standard deviation of the predicted spectra as solely due to the cosmic variance induced error without any sample variance even in presence of partial sky observations. 

\subsection{Significance ratios}
\label{sign_Dl}
\begin{figure}
\centering
\includegraphics[scale=0.38]{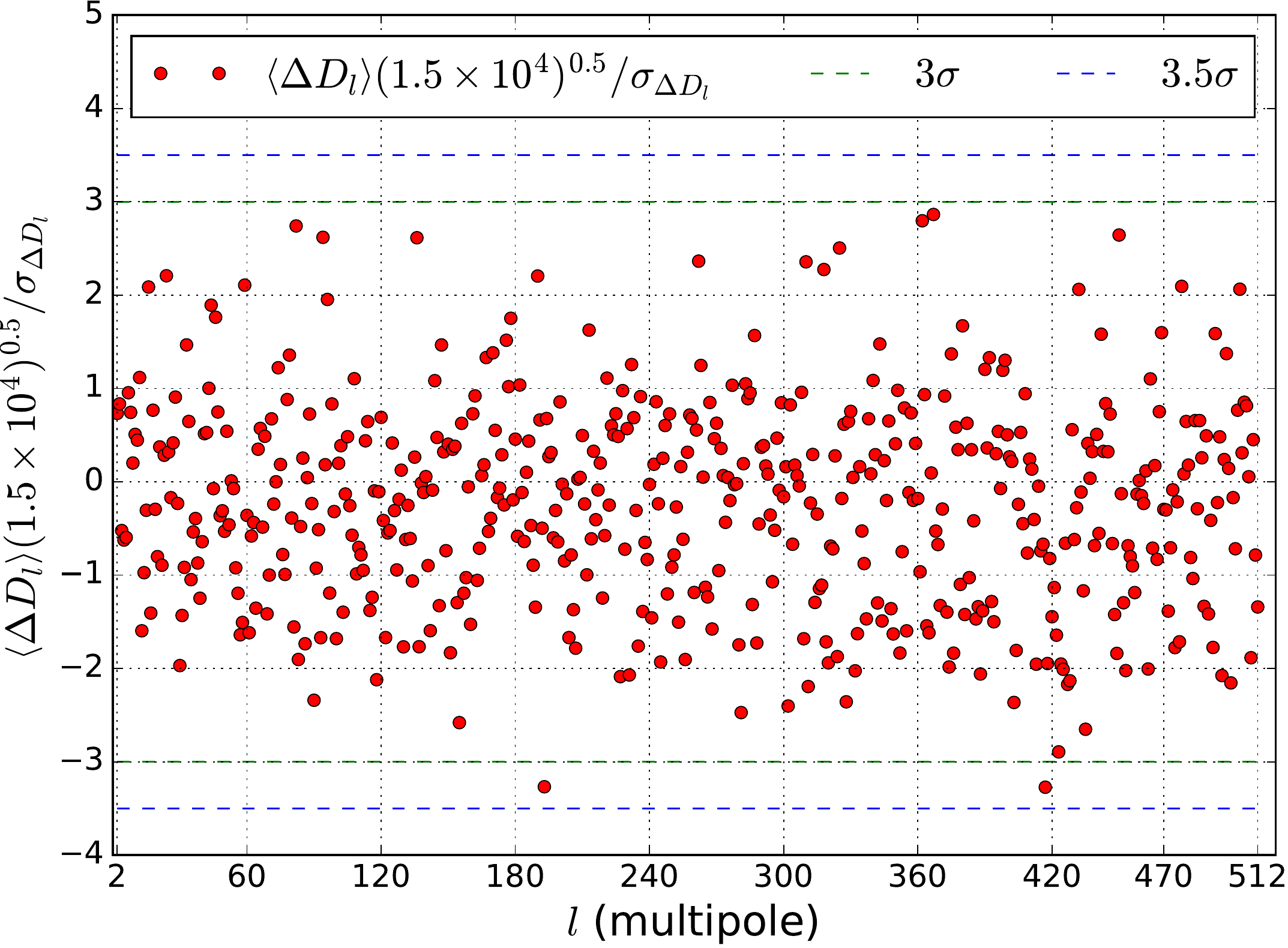}
\caption{Figure shows the significance ratios (red) of the predicted full sky spectrum ($D_l$) for the multipole range $2 \leq l \leq 512$. All the significance ratios (red) are entirely within $3.5\sigma$ (blue dashed line) error interval. Moreover, these ratios (red) are within $3\sigma$ (green dashed line) error interval except for two multipoles, namely, $l=193, 417$. See the section~\ref{sign_Dl} for the detailed discussions corresponding to this figure.}
\label{significance}
\end{figure}
\begin{figure*}
\centering
\includegraphics[scale=0.7]{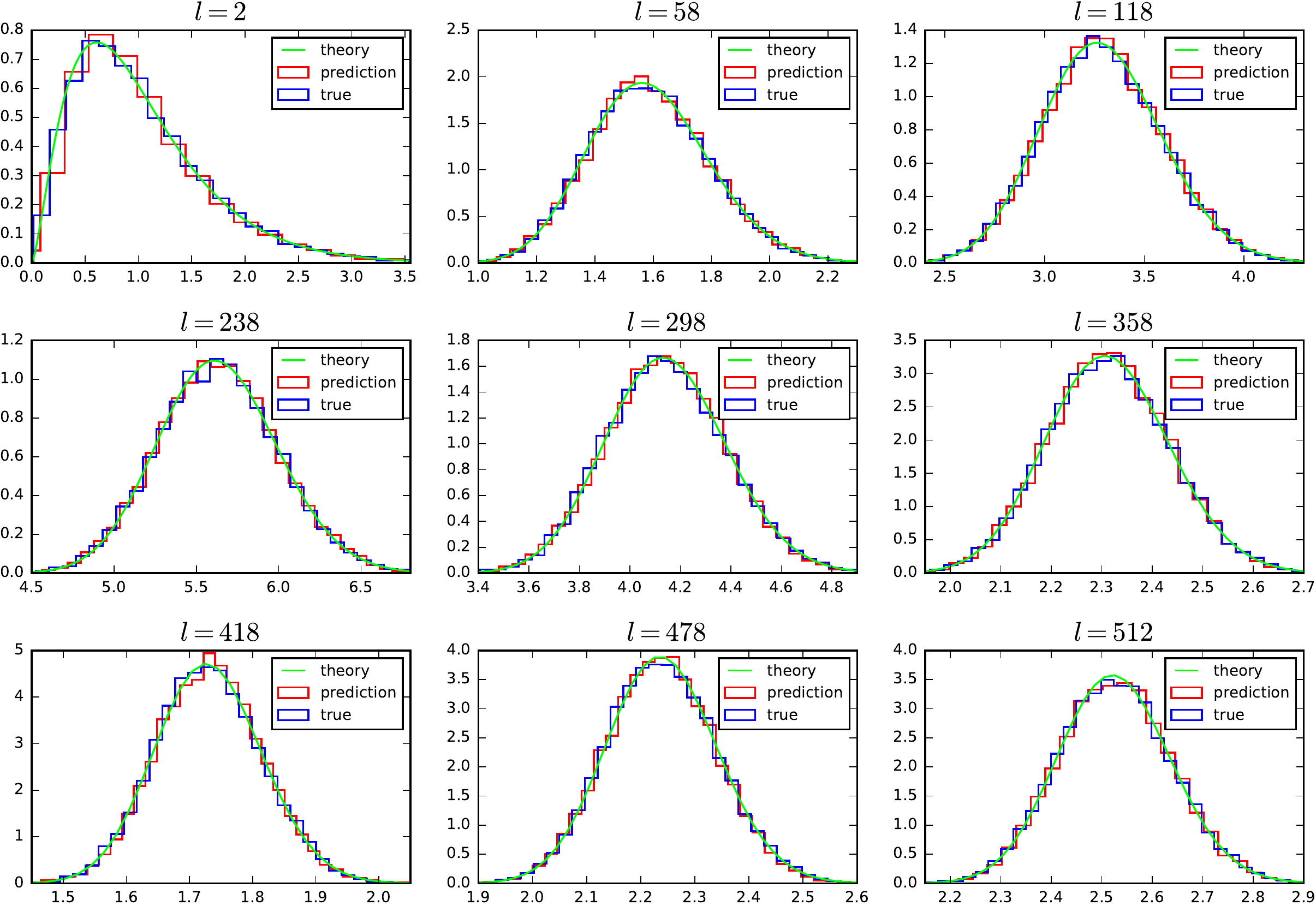}
\caption{Every sub-figure shows the probability density (normalized histogram) of predicted (red) and target (blue) $D_{l} = l(l+1)C_{l}/2\pi$ as well as the conditional probability density (lime) of full sky (given theoretical spectrum) for different multipoles. Predicted (red) probability densities give an excellent agreement with target (blue) and theoretical (lime) probability densities. Horizontal axis of each sub-figure represents full sky $D_{l}$ in $1000 \mu K^{2}$ unit.}
\label{hist}
\end{figure*}
To assess the effectiveness and accuracy of the predictions of the ANN, we estimate the significance ratio of the predicted full sky spectrum $D_{l}$ at each multipole. In Figure~\mbox{\ref{significance}}, we show the scatter of mean predicted spectrum around the mean target spectrum by computing the significance ratios along with several error interval regions. Atfirst, we calculate the differences ($\Delta D_l$) between target and predicted spectra for each realization. Then, we find the mean ($\left<\Delta D_l\right>$) and the standard deviation ($\sigma_{\Delta D_l}$) of these differences. After that, we estimate the standard error of mean (SEM) of these differences from the standard deviation dividing by the square root of the number ($1.5 \times 10^4$) of realizations. Finally, we calculate the significance ratios dividing the mean of those differences by SEM. We can note that the entire significance ratio points (red) are within $3.5\sigma$ (blue dashed line) error interval (more specifically the ratios are within $3\sigma$ (green dashed line) error interval except for two multipoles, namely, $l = 193,417$), which implies the predictions from our ANN system are excellently unbiased. So, we conclude that the \textit{model averaging ensemble}~\citep{Lai_2021} method effectively reduces the epistemic uncertainties in the predictions of our ANN system.

\subsection{Probability density}
\label{PD_Dl}
We verify that the predicted $D_{l}$ at every multipole follows $\chi^{2}$ distributions as we expect from the Gaussian nature of CMB anisotropies. We obtain the probability density of predicted $D_{l}$ as well as the target $D_{l}$ by finding normalized histograms for each multipole. We also calculate the conditional probability densities of full sky CMB angular spectra given theoretical $D_{l}^{th}$~\citep{Sudevan_2020}, for every multipole. We present the normalized histograms in Figure~\ref{hist} for some different multipoles. Horizontal axis of each sub-figure represents the values of full sky $D_{l}$ in $1000\mu K^{2}$ unit. Normalized histograms show that the probability densities of predicted $D_{l}$ agree very well with the probability densities of target $D_{l}$ for each multipole. These normalized histograms also agree with the theoretical probability densities excellently. So the predicted full sky spectrum from our ANN system preserves the statistics of $\chi^{2}$ distribution very nicely.
\subsection{Correlation matrix}
\label{corr_Dl}
\begin{figure*}
\centering
\includegraphics[scale=0.5]{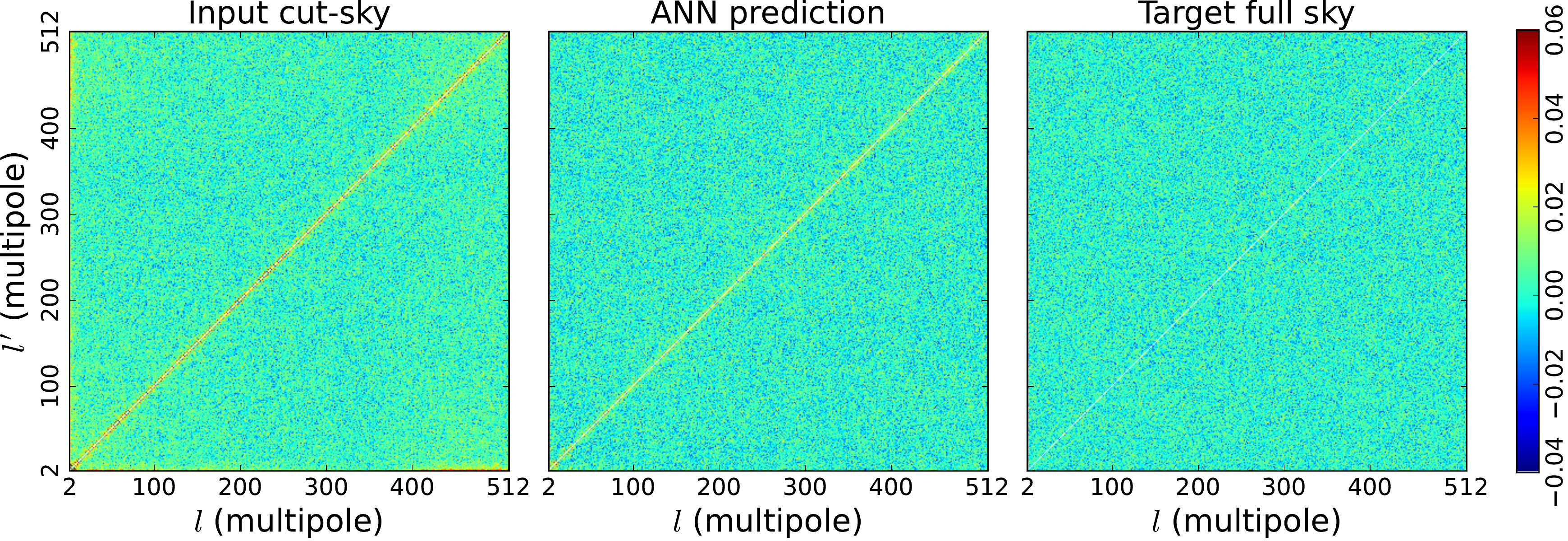}
\caption{Left panel of the figure represents the correlation matrix corresponding to the input cut-sky power spectra for the multipole range $2 \leq l \leq 512$. Middle panel of the figure shows the correlation matrix of the ANN predicted full sky spectra for the same multipole range. Right panel of the figure shows the correlation matrix corresponding to the target full sky spectra for the same multipole range. We clip the colour scale within $-0.04$ and $0.06$ to look into the off-diagonal area of the matrices clearly. As the diagonal elements of these matrices are unity, we mask these diagonal values using white colour to visualize only the off-diagonal area. See the section~\ref{corr_Dl} for the detailed discussion.}
\label{corr}
\end{figure*}
\begin{figure*}
\centering
\includegraphics[scale=0.5]{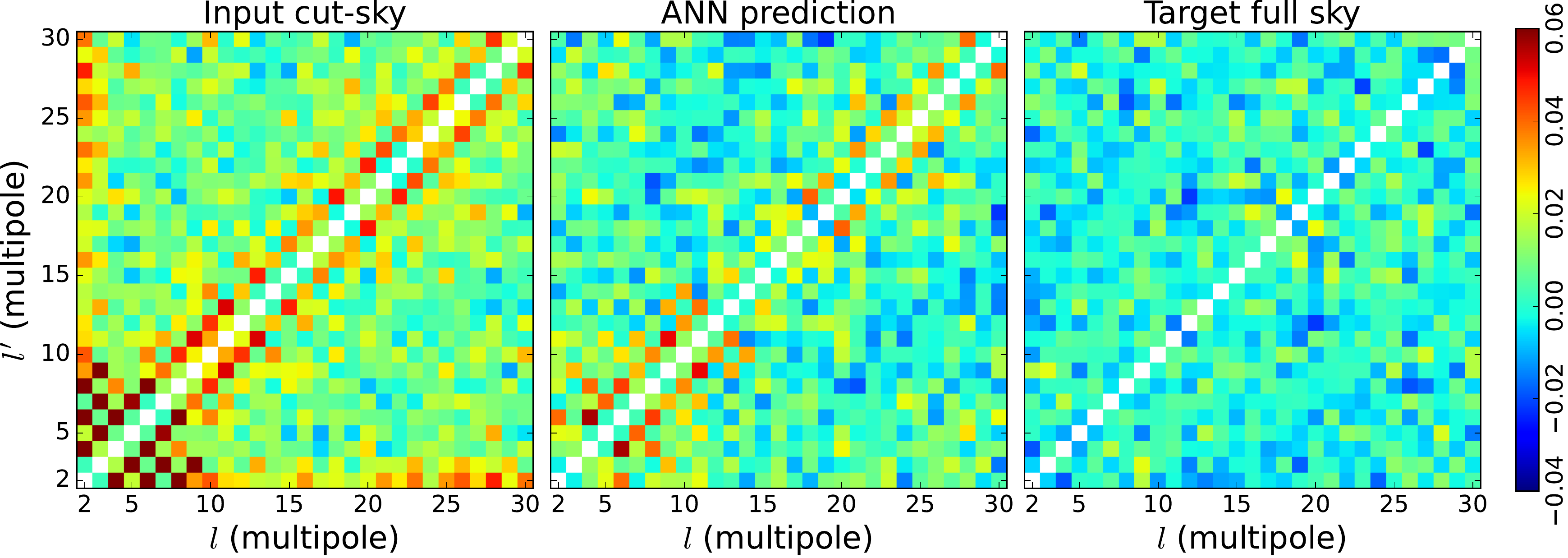}
\caption{Figure shows the zoom in area of the matrices of Figure~\ref{corr} for the multipole range $2 \leq l \leq 30$. Left panel of the figure is corresponding to the input cut-sky spectra, the middle panel represents for the full sky spectra predicted by our ANN system, and the right panel presents for the target full sky spectra. Here, the colour scale is clipped within $-0.04$ and $0.06$ and the diagonal values are masked by using white colour. See the text of section~\ref{corr_Dl} for the detailed discussion.}
\label{corr_spec}
\end{figure*}
We estimate the multipole-space correlation matrix for the predicted full sky spectra for the multipole range $2 \leq l \leq 512$ to investigate any possible correlations that may exist in our predictions. Moreover, we compare the correlation matrix of predicted spectra with the correlation matrix corresponding to input cut-sky spectra and also with the correlation matrix of the target full sky spectra. In Figure~\ref{corr}, we show the multipole-multipole ($l'$-$l$) correlation matrices for input cut-sky spectra, the ANN predicted full sky spectra, and the target full sky spectra for the multipole range $2 \leq l \leq 512$. In this figure, we clip the color scale within $-0.04$ and $0.06$. We also set the diagonal values equal to zero (in place of unity)  to zoom in the off-diagonal area of the matrices and fill these diagonal pixels with white colour.
\begin{table}[h!]
\centering
\caption{Table shows the pixel indices (i.e., $l$ and $l'$) of the correlations larger than $0.0563$ in the input cut-sky spectra.}
\begin{tabular}{c|c||c|c||c|c||c|c}
\hline
$l$ & $l'$ & $l$ & $l'$ & $l$ & $l'$ & $l$ & $l'$\\
\hline\hline
$2$ & $4,6,8,474,493$ & $58$ & $60$ & $197$ & $199$ & $313$ & $315$\\
$3$ & $5,7,9$ & $91$ & $93$ & $222$ & $224$ & $435$ & $437$\\
$4$ & $6$ & $106$ & $108$ & $225$ & $227$ & $461$ & $463$\\
$5$ & $7$ & $134$ & $136$ & $231$ & $233$ & $476$ & $478$\\
$6$ & $8$ & $179$ & $181$ & $239$ & $241$ & & \\
\hline
\end{tabular}
\label{corr_tab}
\end{table}
\begin{figure*}
\centering
\includegraphics[scale=0.7]{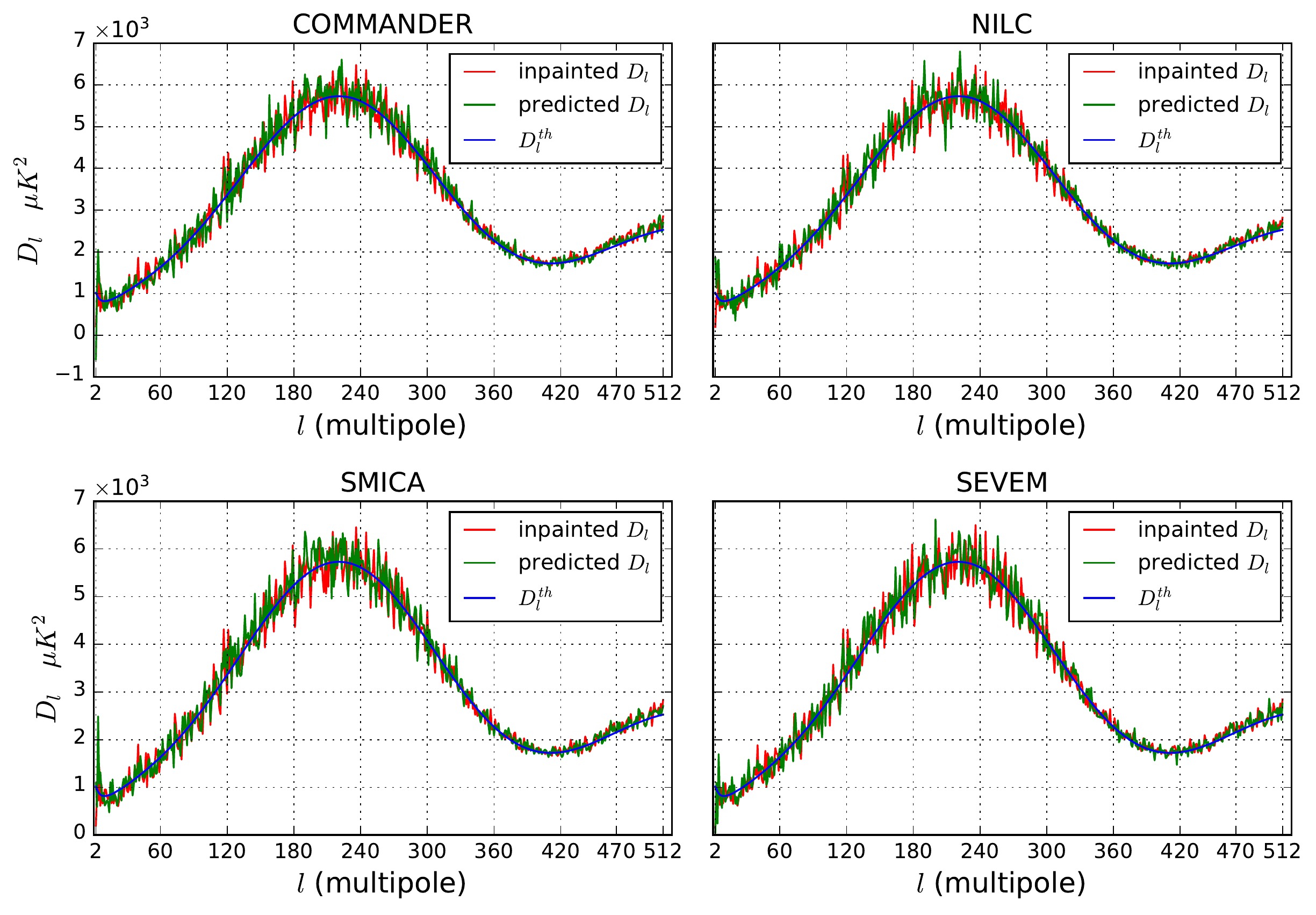}
\caption{Figure shows the excellent agreement of the predicted (green) $D_{l} = l(l+1)C_{l}/2\pi$ and inpainted (red) $D_{l}$, for COMMANDER, NILC, SMICA and SEVEM, in the multipole range $2\leq l \leq 512$. Predicted (green) spectra also agree with the theoretical (blue) spectrum excellently shown in the same graph.}
\label{real_data}
\end{figure*}
We note that the off-diagonal correlations corresponding to our ANN system are in between $-0.0347$ and $0.0563$, and the off-diagonal correlations of input cut-sky spectra are within $-0.0322$ and $0.1295$, where the off-diagonal correlations of the target full sky spectra are within $-0.036$ and $0.0339$. The pixel indices (i.e., $l$ and $l'$) of the correlation matrix corresponding to input cut-sky spectra, where the correlations are larger than $0.0563$, are presented in the Table~\ref{corr_tab}. In Figure~\ref{corr_spec}, we show a specific off-diagonal region for the multipole range $2 \leq l \leq 30$ to compare the larger correlations of input cut-sky with the correlations corresponding to ANN system as well as with the same of the target full sky spectra at these same points. These correlations of input cut-sky spectra are effectively diminished in the predicted spectra using our ANN method.

\subsection{Predictions for observed CMB maps}
\label{predict_real_data}
We further apply our ANN system, trained with $1.2 \times 10^{5}$ training samples, on the available unpainted foreground cleaned CMB maps. We collect these CMB maps (e.g., COMMANDER, NILC, SMICA and SEVEM)\footnote{\url{https://pla.esac.esa.int/}} from~\cite{Planck4_2020}. We apply Kq85-mask, for $N_{side}=256$, on these four unpainted full sky maps. We obtain unpainted partial sky spectra, using $\texttt{healpy.sphtfunc.anafast}$, given these four unpainted partial sky maps for maximum multipole $l_{max}=767$. We use these unpainted partial sky $\tilde{D}_{l}$ as input data to predict corresponding full sky $D_{l}$, up to maximum multipole $l_{max}=512$, using our ANN architecture.

We use inpainted full sky spectra for comparing the consistency of predictions of our ANN system on Planck data~\citep{Planck4_2020}. These inpainted spectra are obtained from four inpainted CMB intensity maps (e.g., COMMANDER, NILC, SMICA and SEVEM), provided by Planck mission. Inpainted maps are generated by~\cite{Planck4_2020} from unpainted foreground cleaned CMB maps replacing some pixels of galactic plane, very bright sources and extragalactic point sources regions by a sample realization drawn from the posterior distribution given the un-masked pixels. The method preserves the statistical properties of the entire even after infilling some regions of the sky. Interestingly, the predictions from our exercise, shown in Figure~\ref{real_data}, exhibit an excellent agreement with theoretical spectrum as well as the inpainted full sky $D_{l}$ obtained from inpainted CMB maps~\citep{Planck4_2020}.

\section{Discussions and Conclusions}
\label{conclusion}
Accurate estimation of full sky CMB angular power spectrum ($C_{l}$), from CMB map after masking off some finite regions, may be a problem of utmost importance for reliable extraction of cosmological information from the observed CMB maps. For large pixel array data, as is the case for high resolution CMB maps, the available methods can be computationally expensive (e.g., the exact maximum likelihood approach) or may not be optimal (e.g, over large angular scales and limited by sample variance). In such cases, training an artificially intelligent system seems to be an effective alternative method which can produce optimal results with the requirements of only a small amount of available computing resources. In the current work, we evolve the earlier work of~\cite{Chanda_2021} for $N_{side} = 16$, to predict the full sky CMB angular power spectrum using ANN after application of Kq85-Mask using $N_{side} = 256$ HEALPix maps and up to $l_{max} = 512$. The mask removes both the (strongly contaminated) galactic plane and the positions of extragalactic bright point sources as required by a high resolution CMB analysis.

There are two major advantages of using the ANN for estimating the full sky CMB angular power spectrum from the partial sky CMB maps. These are summarized below.\\
\hspace*{5pt}(i) First, our method can produce unbiased estimate of the underlying full sky CMB power spectrum over the entire multipole range $2 \leq l \leq 512$ even in presence of missing data. In the ANN method of our work, the ensemble average of $1.5 \times 10^4$ number of predicted spectra agrees excellently with the ensemble average of corresponding target spectra and also with the theoretical spectrum at each multipole.\\
\hspace*{5pt}(ii) Secondly, our method does not suffer from the issues of sample variance due to loss of modes over the partial sky from which the full sky predictions of CMB spectra are desired. Therefore, statistically our method can produce an ensemble of spectra with only cosmic variance induced error without the sample variance which plagues the classical approaches. Recovery of full sky estimates of the CMB spectra from the masked maps without sampling variance (and without the missing data or any other independent additional data) is indeed a very appealing and desirable property of the ANN system. Bypassing the sample variances in our method can be explained by noting that we are training the intelligent system to learn the unknown mapping which exists in between the partial sky and the corresponding full sky angular power spectra. Hence, our method should not be treated as a traditional approach of recovering the full sky estimate of the angular power spectrum from the partial sky maps which leave to the sample variances in the recovered spectrum. Our method is fundamentally and functionally different from the reconstruction of the power spectrum from the partial sky maps where the sample variances inevitably contribute.

In the foregoing paragraphs, we have discussed two major advantages to utilize the ANN system to predict full sky spectrum from partial sky CMB map. The fundamental reason behind both the advantages is that following ANN approach we predict the full sky spectra based upon an intelligent system  which is capable of learning the mapping between the input and targets from the training phase. Since the learning has been satisfactory, the intelligent system well preserves all the statistical properties of the target variables that it sees during training and consequently encodes their informations into its perception stored into several matrix images  of weights and biases. Due to this reason, the predicted spectra by the ANN system does not show any significant correlation like the target spectra and the predictions does not suffer from any problem of sample variances. Due to the same reason we also obtain the third important advantage of the ANN approach, namely the distributions of predicted spectra and target spectra agree with each other with a very good accuracy. This opens up an interesting possibility of using the outcome of the ANN prediction into the likelihood analysis of the cosmological parameters. The improved error estimates on the full sky prediction of the angular spectra from the partial sky maps and statistical similarity of the predictions with the target spectra  are expected to lead to improved error estimates on the derived cosmological parameters. We also note that we employ model averaging method to reduce the epistemic uncertainty in our prediction which represents any possible bias (systematic error) in the predictions. In a future article we explore these problems in detail.

In our analysis the testing and training sets are computed from the same theoretical CMB angular spectrum, although in reality the theoretical spectrum is not known apriori. If the fiducial spectrum used by us for training is reasonably close to the actual underlying theoretical spectrum, any bias in our ANN predictions due to incorrect knowledge of the theoretical spectrum is expected to be small. A possible future method to improve the current work will be to predict an estimate of the full sky theoretical spectrum simultaneously with the observed full sky spectrum given the (observed) partial sky spectrum. We will explore this project in a future publication.

Moreover, the relation (mapping) between CMB angular power spectrum estimated from the partial sky and the corresponding full sky spectrum depends upon the mask. A different mask is expected to lead to a different mapping function between the two (e.g., the mode-mode coupling matrix in the usual MASTER approach depends on the angular power spectrum of the mask). Hence, for a different mask than what we used in the current article we would need to train the ANN system anew for the input spectra computed from the partial sky maps obtained by using the new mask. We note in passing that for this new mask we do not need to construct a new ANN system. We merely train the original ANN with the new partial sky spectra to predict the full sky spectra from the correponding unknown partial sky spectra.
 
The ANN predicts both the full sky spectra and the corresponding cosmic variance induced error at each multipole. We use \textit{model averaging ensemble} methods to reduce the epistemic error. Using a sufficiently generous mask in our analysis, we find that the mean of the predicted full sky spectra agrees within approximately $3\sigma$ ($99.7\%$ confidence level) with the underlying theoretical spectrum indicating the predictions are unbiased. On applying our trained ANN on the partial sky (Kq85 masked) maps generated from unpainted forground cleaned CMB maps, COMMANDER, NILC, SMICA and SEVEM, we find an excellent agreement of predicted spectra with the corresponding full sky spectra from cleaned (inpainted) maps~\citep{Planck4_2020}. This again shows that our ANN has learnt satisfactorily to reconstruct the lost information due to masking reliably based upon the training. In a future article we will utilize ANN in the problem of CMB component separation over partial sky so as to reconstruct the joint distributions of (partial sky) cleaned CMB map and the corresponding full sky spectrum.

\section*{Acknowledgements}
Authors acknowledge the use of the open-source packages HEALPix\footnote{\url{https://healpix.sourceforge.io/}},  TensorFlow\footnote{\url{https://www.tensorflow.org/}} and $\texttt{Google Colab}$\footnote{\url{https://colab.research.google.com/?utm_source=scs-index}}.


\begin{thebibliography}{}
\makeatletter
\relax
\def\mn@urlcharsother{\let\do\@makeother \do\$\do\&\do\#\do\^\do\_\do\%\do\~}
\def\mn@doi{\begingroup\mn@urlcharsother \@ifnextchar [ {\mn@doi@}
  {\mn@doi@[]}}
\def\mn@doi@[#1]#2{\def\@tempa{#1}\ifx\@tempa\@empty \href
  {http://dx.doi.org/#2} {doi:#2}\else \href {http://dx.doi.org/#2} {#1}\fi
  \endgroup}
\def\mn@eprint#1#2{\mn@eprint@#1:#2::\@nil}
\def\mn@eprint@arXiv#1{\href {http://arxiv.org/abs/#1} {{\tt arXiv:#1}}}
\def\mn@eprint@dblp#1{\href {http://dblp.uni-trier.de/rec/bibtex/#1.xml}
  {dblp:#1}}
\def\mn@eprint@#1:#2:#3:#4\@nil{\def\@tempa {#1}\def\@tempb {#2}\def\@tempc
  {#3}\ifx \@tempc \@empty \let \@tempc \@tempb \let \@tempb \@tempa \fi \ifx
  \@tempb \@empty \def\@tempb {arXiv}\fi \@ifundefined
  {mn@eprint@\@tempb}{\@tempb:\@tempc}{\expandafter \expandafter \csname
  mn@eprint@\@tempb\endcsname \expandafter{\@tempc}}}
  
\bibitem[\protect\citeauthoryear{{Abadi} et al.}{2015}]{Abadi_2015}
Abadi M., et al., 2015, TensorFlow: Large-Scale Machine Learning on Heterogeneous Systems,
\href{http://download.tensorflow.org/paper/whitepaper2015.pdf}{download.tensorflow.org/paper/whitepaper2015.pdf}

\bibitem[\protect\citeauthoryear{{Acquaviva} et al.}{2003}]{Acquaviva_2003}
Acquaviva, V., Bartolo, N., Matarrese, S. and Riotto, A., 2003, \mn@doi [Nuclear Physics B]
{https://doi.org/10.1016/S0550-3213(03)00550-9}, \href
{https://www.sciencedirect.com/science/article/pii/S0550321303005509}{667, 119}

\bibitem[\protect\citeauthoryear{{Allen} et al.}{1987}]{Allen_1987}
Allen, T., J., Grinstein, B., and Wise, M., B., 1987, \mn@doi [Physics Letters B]
{https://doi.org/10.1016/0370-2693(87)90343-1}, \href
{https://www.sciencedirect.com/science/article/pii/0370269387903431}{197, 66}

\bibitem[\protect\citeauthoryear{{Alsing} et al.}{2015}]{Alsing_2015}
Alsing J., Heavens A., Jaffe A. H., Kiessling A., Wandelt B., Hoffmann T., 2015, \mn@doi [\mnras]
{10.1093/mnras/stv2501}, \href
{https://doi.org/10.1093/mnras/stv2501}{455, 4452}

\bibitem[\protect\citeauthoryear{{Baccigalupi} et al.}{2000}]{Baccigalupi_2000}
Baccigalupi C., Bedini L., Burigana C., de Zotti G., Farusi A., Maino D., Maris M., Perrotta F., Salerno E., Toffolatti L. and Tonazzini A., 2000, \mn@doi [\mnras]
{10.1046/j.1365-8711.2000.03751.x}, \href
{https://doi.org/10.1046/j.1365-8711.2000.03751.x}{318, 769}

\bibitem[\protect\citeauthoryear{{Bennett} et al.}{1996}]{Bennett_1996}
Bennett C. L. et al., 1996, \mn@doi [\apj]
{10.1086/310075}, \href
{https://doi.org/10.1086/310075}{464, L1}

\bibitem[\protect\citeauthoryear{{Bennett} et al.}{2013}]{Bennett_2013}
Bennett C. L. et al., 2013, \mn@doi [\apjs]
{10.1088/0067-0049/208/2/20}, \href
{https://doi.org/10.1088\%2F0067-0049\%2F208\%2F2\%2F20}{208, 20}

\bibitem[\protect\citeauthoryear{{Bond} et al.}{1998}]{Bond_1998}
Bond J. R., Jaffe A. H., Knox L., 1998, \mn@doi [\prd]
{10.1103/PhysRevD.57.2117}, \href
{https://link.aps.org/doi/10.1103/PhysRevD.57.2117}{57, 2117}

\bibitem[\protect\citeauthoryear{{Chanda} \& {Saha}}{2021}]{Chanda_2021}
Chanda P. and Saha R. 2021, \mn@doi [\mnras]
{10.1093/mnras/stab2753}, \href
{https://doi.org/10.1093/mnras/stab2753}{508, 4600}

\bibitem[\protect\citeauthoryear{{Dialektopoulos} et al.}{2021}]{Dialektopoulos_2021}
Dialektopoulos K., Said J. L., Mifsud J., Sultana J. and Adami K. Z., 2021,
\href{https://arxiv.org/abs/2111.11462}{arxiv.org/abs/2111.11462}

\bibitem[\protect\citeauthoryear{{Elsner} et al.}{2016}]{Elsner_2016}
Elsner F., Leistedt B., Peiris H. V., 2016, \mn@doi [\mnras]
{10.1093/mnras/stw2752}, \href
{https://doi.org/10.1093/mnras/stw2752}{465, 1847}

\bibitem[\protect\citeauthoryear{{Eriksen} et al.}{2004}]{Eriksen_2004}
Eriksen H. K., O'Dwyer I. J., Jewell J. B., Wandelt B. D., Larson D. L., Gorski K. M., Levin S., Banday A. J. and Lilje P. B., 2004, \mn@doi [\apjs]
{10.1086/425219}, \href
{https://doi.org/10.1086/425219}{155, 227}

\bibitem[\protect\citeauthoryear{{Escamilla-Rivera} et al.}{2020}]{Escamilla_2020}
Escamilla-Rivera C., Quintero M. A. C. and Capozziello S., 2020, \mn@doi [\jcap]
{10.1088/1475-7516/2020/03/008}, \href
{https://doi.org/10.1088/1475-7516/2020/03/008}{2020, 008}

\bibitem[\protect\citeauthoryear{{Falk} et al.}{1992}]{Falk_1992}
Falk, T., Rangarajan, R. and Srednicki, M., 1992, \mn@doi [\prd]
{10.1103/PhysRevD.46.4232}, \href
{https://link.aps.org/doi/10.1103/PhysRevD.46.4232}{46, 4232}

\bibitem[\protect\citeauthoryear{{Fixsen} et al.}{1996}]{Fixsen_1996}
Fixsen D. J., Cheng E. S., Gales J. M., Mather J. C., Shafer R. A. and Wright E. L., 1996, \mn@doi [\apj]
{10.1086/178173}, \href
{https://doi.org/10.1086/178173}{473, 576}

\bibitem[\protect\citeauthoryear{{Gal} et al.}{2017}]{Gal_2017}
Gal Y., Hron J., Kendall A., 2017, Concrete Dropout,
\href{https://arxiv.org/abs/1705.07832}{arxiv.org/abs/1705.07832}

\bibitem[\protect\citeauthoryear{{Gangui} et al.}{1994}]{Gangui_1994}
{Gangui}, A., {Lucchin}, F., {Matarrese}, S. and {Mollerach}, S., 1994, \mn@doi [\apj]
{10.1086/174421}, \href
{https://ui.adsabs.harvard.edu/abs/1994ApJ...430..447G}{430, 447}

\bibitem[\protect\citeauthoryear{{G{\'{o}}mez-Vargas}, {Esquivel} et al.}{2021}]{Gomez1_2021}
G{\'{o}}mez-Vargas I., Esquivel R. M., Garc{\'{\i}}a-Salcedo R. and V{\'{a}}zquez J. A., 2021, \mn@doi [Journal of Physics: Conference Series]
{10.1088/1742-6596/1723/1/012022}, \href
{https://doi.org/10.1088/1742-6596/1723/1/012022}{1723, 012022}

\bibitem[\protect\citeauthoryear{{G{\'{o}}mez-Vargas}, {Vázquez} et al.}{2021}]{Gomez2_2021}
G{\'{o}}mez-Vargas I., Vázquez J. A., Esquivel R. M. and García-Salcedo R., 2021, 
\href{https://arxiv.org/abs/2104.00595}{arxiv.org/abs/2104.00595}

\bibitem[\protect\citeauthoryear{{Gorski}}{1994}]{Gorski_1994}
Gorski M. K., 1994, \mn@doi [\apjl]
{10.1086/187444}, \href
{https://ui.adsabs.harvard.edu/abs/1994ApJ...430L..85G}{430, L85}

\bibitem[\protect\citeauthoryear{{Gorski} et al.}{1994}]{Gorski2_1994}
{Gorski}, K.~M., {Hinshaw}, G., {Banday}, A.~J., {Bennett}, C.~L., {Wright}, E.~L., {Kogut}, A., {Smoot}, G.~F. and {Lubin}, P., 1994, \mn@doi [\apjl]
{10.1086/187445}, \href
{https://ui.adsabs.harvard.edu/abs/1994ApJ...430L..89G}{430, L89}

\bibitem[\protect\citeauthoryear{{Gorski} et al.}{1996}]{Gorski_1996}
G{\'{o}}rski K. M., Banday A. J., Bennett C. L., Hinshaw G., Kogut A.,Smoot G. F.and Wright E. L., 1996, \mn@doi [\apjl]
{10.1086/310077}, \href
{https://doi.org/10.1086/310077}{464, L11}

\bibitem[\protect\citeauthoryear{{Gorski}}{1997}]{Gorski_1997}
Gorski M. K., 1997,
\href
{https://arxiv.org/abs/astro-ph/9701191}{arXiv:astro-ph/9701191}

\bibitem[\protect\citeauthoryear{{Gorski} et al.}{2005}]{Gorski_2005}
Gorski M. K., Hivon E., Banday J. A., Wandelt D. B., Hansen K. F., Reinecke M. and Bartelmann M., 2005, \mn@doi [\apj]
{10.1086/427976}, \href
{https://doi.org/10.1086/427976}{622, 759}

\bibitem[\protect\citeauthoryear{{Graff} et al.}{2012}]{Graff_2012}
Graff P., Feroz F., Hobson M. P. and Lasenby A., 2012, \mn@doi [\mnras]
{10.1111/j.1365-2966.2011.20288.x}, \href
{https://doi.org/10.1111/j.1365-2966.2011.20288.x}{421, 169}

\bibitem[\protect\citeauthoryear{{Guth} \& {Pi}}{1982}]{Guth_1982}
Guth, Alan H. and Pi, So-Young, 1982, \mn@doi [\prl]
{10.1103/PhysRevLett.49.1110}, \href
{https://link.aps.org/doi/10.1103/PhysRevLett.49.1110}{49, 1110}

\bibitem[\protect\citeauthoryear{{Hajian} \& {Souradeep}}{2004}]{Hajian_2004}
Hajian, A. and Souradeep, T., 2004, 
\href{https://arxiv.org/pdf/astro-ph/0501001.pdf}{arxiv.org/pdf/astro-ph/0501001.pdf}

\bibitem[\protect\citeauthoryear{{Hanany} et al.}{2019}]{Hanany_2019}
Hanany, S. et al., 2019, 
\href{https://arxiv.org/pdf/1902.10541.pdf}{arxiv.org/pdf/1902.10541.pdf}

\bibitem[\protect\citeauthoryear{{Hansen} et al.}{2002}]{Hansen_2002}
Hansen F. K., Gorski K. M., Hivon E., 2002, \mn@doi [\mnras]
{10.1046/j.1365-8711.2002.05878.x}, \href
{https://doi.org/10.1046/j.1365-8711.2002.05878.x}{336, 1304}

\bibitem[\protect\citeauthoryear{{Hazumi} et al.}{2020}]{Hazumi_2020}
Hazumi, M. et al., 2020,
\href{https://arxiv.org/pdf/2101.12449.pdf}{arxiv.org/pdf/2101.12449.pdf}

\bibitem[\protect\citeauthoryear{{Hecht-Nielsen}}{1992}]{Hecht-Nielsen_1992}
Hecht-Nielsen R., 1992, in , Neural Networks for Perception. Academic Press, pp 65–93,
\href{https://doi.org/10.1016/B978-0-12-741252-8.50010-8}{doi.org/10.1016/B978-0-12-741252-8.50010-8}

\bibitem[\protect\citeauthoryear{{Hinshaw} et al.}{2013}]{Hinshaw_2013}
Hinshaw G. et al., 2013, \mn@doi [\apjs]
{10.1088/0067-0049/208/2/19}, \href
{https://doi.org/10.1088/0067-0049/208/2/19}{208, 19}

\bibitem[\protect\citeauthoryear{{Hinton} et al.}{2012}]{Hinton_2012}
Hinton G., Srivastava N., Swersky K., 2012, rmsprop: Divide the gradient by a running average of its recent magnitude, Neural networks for machine learning—Lecture 6
\href{https://www.cs.toronto.edu/~hinton/coursera/lecture6/lec6.pdf}{cs.toronto.edu/hinton/coursera/lecture6/lec6.pdf}
  
\bibitem[\protect\citeauthoryear{{Hivon} et al.}{2002}]{Hivon_2002}
Hivon E., Gorski K. M., Netterfield C. B., Crill B. P., Prunet S. and Hansen F., 2002, \mn@doi [\apj]
{10.1086/338126}, \href
{https://doi.org/10.1086/338126}{567, 2}

\bibitem[\protect\citeauthoryear{{Hornik}}{1991}]{Hornik_1991}
Hornik K., 1991, \mn@doi [Neural Networks]
{https://doi.org/10.1016/0893-6080(91)90009-T}, \href
{https://www.sciencedirect.com/science/article/pii/089360809190009T}{4, 251}

\bibitem[\protect\citeauthoryear{{Hortua} et al.}{2020}]{Hortua_2020}
Hortua H. J., Volpi R., Marinelli D. and Malago L., 2020,
\href{https://arxiv.org/abs/2011.14276}{arxiv.org/abs/2011.14276}

\bibitem[\protect\citeauthoryear{{Hou} et al.}{2014}]{Hou_2014}
Hou Z. et al., 2014, \mn@doi [\apj]
{10.1088/0004-637x/782/2/74}, \href
{https://doi.org/10.1088/0004-637x/782/2/74}{782, 74}

\bibitem[\protect\citeauthoryear{{Kendall} {\&} {Gal}}{2017}]{Kendall_2017}
Kendall A., Gal Y., 2017, What Uncertainties Do We Need in Bayesian Deep Learning for Computer Vision?
\href{https://arxiv.org/abs/1703.04977.pdf}{arxiv.org/abs/1703.04977.pdf}

\bibitem[\protect\citeauthoryear{{Kingma} {\&} {Ba}}{2014}]{Kingma_2014}
Kingma D. P., Ba J., 2014, Adam: A method for stochastic optimization
\href{https://arxiv.org/abs/1412.6980}{arxiv.org/abs/1412.6980}

\bibitem[\protect\citeauthoryear{{Lai} et al.}{2021}]{Lai_2021}
Lai Y., Shi Y., Han Y., Shao Y., Qi M. and Li B., 2021,
\href{https://arxiv.org/abs/2104.12953}{arxiv.org/abs/2104.12953}

\bibitem[\protect\citeauthoryear{{Maldacena}}{2003}]{Maldacena_2003}
Maldacena, J., 2003, \mn@doi [Journal of High Energy Physics]
{10.1088/1126-6708/2003/05/013}, \href
{https://doi.org/10.1088/1126-6708/2003/05/013}{2003,013}

\bibitem[\protect\citeauthoryear{{Mancini} et al.}{2022}]{Mancini_2022}
Mancini A. S., Piras D., Alsing J., Joachimi B. and Hobson M. P., 2022, \mn@doi [\mnras]
{10.1093/mnras/stac064}, \href
{https://doi.org/10.1093/mnras/stac064}{1365-2966}

\bibitem[\protect\citeauthoryear{{Mather} et al.}{1999}]{Mather_1999}
Mather J. C., Fixsen D. J., Shafer R. A., Mosier C. and Wilkinson D. T., 1999, \mn@doi [\apj]
{10.1086/306805}, \href
{https://doi.org/10.1086/306805}{512, 511}

\bibitem[\protect\citeauthoryear{{Moss}}{2020}]{Moss_2020}
Moss A., 2020, \mn@doi [\mnras]
{10.1093/mnras/staa1469}, \href
{https://doi.org/10.1093/mnras/staa1469}{496, 328}

\bibitem[\protect\citeauthoryear{{Netterfield} et al.}{1997}]{Netterfield_1997}
{Netterfield}, C.~B., {Devlin}, M.~J., {Jarosik}, N., {Page}, L. and {Wollack}, E.~J., 1997, \mn@doi [\apj]
{10.1086/303438}, \href
{https://ui.adsabs.harvard.edu/abs/1997ApJ...474...47N}{474, 47}

\bibitem[\protect\citeauthoryear{{Olvera}, {Gómez-Vargas} \& {Vázquez}}{2021}]{Olvera_2021}
Olvera J. de D. R., Gómez-Vargas I. and Vázquez J. A., 2021,
\href{https://arxiv.org/abs/2112.12645}{arxiv.org/abs/2112.12645}

\bibitem[\protect\citeauthoryear{{Peebles}}{1973}]{Peebles_1973}
{Peebles}, P.~J.~E., 1973, \mn@doi [\apj]
{10.1086/152431}, \href
{https://ui.adsabs.harvard.edu/abs/1973ApJ...185..413P}{185, 413}

\bibitem[\protect\citeauthoryear{{Penzias} \& {Wilson}}{1965}]{Penzias_1965}
{Penzias}, A.~A. and {Wilson}, R.~W. 1965, \mn@doi [\apj]
{10.1086/148307}, \href
{https://ui.adsabs.harvard.edu/abs/1965ApJ...142..419P}{142, 419}

\bibitem[\protect\citeauthoryear{{Petroff} et al.}{2020}]{Petroff_2020}
Petroff M. A., Addison G. E., Bennett C. L., Weiland J. L., 2020, \mn@doi [\apj]
{10.3847/1538-4357/abb9a7}, \href
{https://doi.org/10.3847/1538-4357/abb9a7}{903, 104}

\bibitem[\protect\citeauthoryear{{Pinkus}}{1999}]{Pinkus_1999}
Pinkus A., 1999, \mn@doi [Acta Numerica]
{10.1017/S0962492900002919}, \href
{https://doi.org/10.1017/S0962492900002919}{8, 143}

\bibitem[\protect\citeauthoryear{{Planck} {Collaboration} {IV}}{2020}]{Planck4_2020}
Planck Collaboration IV, 2020, \mn@doi [A \& A]
{10.1051/0004-6361/201833881}, \href
{http://dx.doi.org/10.1051/0004-6361/201833881}{641, A4}

\bibitem[\protect\citeauthoryear{{Planck} {Collaboration} {VI}}{2020}]{Planck_2020}
Planck Collaboration VI, 2020, \mn@doi [A \& A]
{10.1051/0004-6361/201833910}, \href
{https://doi.org/10.1051/0004-6361/201833910}{641, A6}

\bibitem[\protect\citeauthoryear{{Reinecke} et al.}{2013}]{Reinecke_2013}
Reinecke M., Seljebotn D. S., 2013, \mn@doi [A \& A]
{10.1051/0004-6361/201321494}, \href
{https://doi.org/10.1051/0004-6361/201321494}{554, A112}

\bibitem[\protect\citeauthoryear{{Ruder}}{2016}]{Ruder_2016}
Ruder S., 2016, An overview of gradient descent optimization algorithms,
\href{http://arxiv.org/abs/1609.04747}{arxiv.org/abs/1609.04747}

\bibitem[\protect\citeauthoryear{{Sievers} et al.}{2013}]{Sievers_2013}
Sievers J. L. et al., 2013, \mn@doi [\jcap]
{10.1088/1475-7516/2013/10/060}, \href
{https://doi.org/10.1088/1475-7516/2013/10/060}{2013, 060}

\bibitem[\protect\citeauthoryear{{Smoot} et al.}{1991}]{Smoot_1991}
{Smoot}, G.~F., {Bennett}, C.~L., {Kogut}, A. et al., 1991, \mn@doi [\apjl]
{10.1086/185988}, \href
{https://ui.adsabs.harvard.edu/abs/1991ApJ...371L...1S}{371, L1}

\bibitem[\protect\citeauthoryear{{Stacey} et al.}{2018}]{Stacey_2018}
Stacey J. G. et al., 2018, \mn@doi [International Society for Optics and Photonics]
{10.1117/12.2314031}, \href
{https://doi.org/10.1117/12.2314031}{10700, 482}

\bibitem[\protect\citeauthoryear{{Sudevan} \& {Saha}}{2020}]{Sudevan_2020}
Sudevan V., Saha R., 2020, \mn@doi [\apj]
{10.3847/1538-4357/ab964e}, \href
{https://doi.org/10.3847/1538-4357/ab964e}{497, 30}

\bibitem[\protect\citeauthoryear{{Sun} et al.}{2020}]{Sun_2020}
Sun S., Cao Z., Zhu H., Zhao J., 2020, \mn@doi [IEEE Transactions on Cybernetics]
{10.1109/TCYB.2019.2950779}, \href
{https://ieeexplore.ieee.org/document/8903465}{50, 3668}

\bibitem[\protect\citeauthoryear{{Wandelt} et al.}{2001}]{Wandelt_2001}
Wandelt, B. D., Hivon, E. and G\'orski, K. M., 2001, \mn@doi [\prd]
{10.1103/PhysRevD.64.083003}, \href
{https://link.aps.org/doi/10.1103/PhysRevD.64.083003}{64, 083003}

\bibitem[\protect\citeauthoryear{{Wandelt} \& {Hansen}}{2003}]{Wandelt_2003}
Wandelt B. D. and Hansen F. K., 2003, \mn@doi [\prd]
{10.1103/PhysRevD.67.023001}, \href
{https://link.aps.org/doi/10.1103/PhysRevD.67.023001}{67, 023001}

\bibitem[\protect\citeauthoryear{{Wang} et al.}{2020}]{Wang_2020}
Wang G. J., Ma X. J., Li S. Y. and Xia J. Q., 2020, \mn@doi [\apjs]
{10.3847/1538-4365/ab620b}, \href
{https://doi.org/10.3847/1538-4365/ab620b}{246, 13}

\makeatother  
\end{thebibliography}
\end{document}